%% file: ms.tex
\documentclass[iop]{emulateapj}

\usepackage{amsmath,amssymb,latexsym,graphics,graphicx,longtable,apjfonts}

\shorttitle{Stellar Masses of SCUBA Galaxies}
\shortauthors{Hainline et al.}

\newcommand{\kms}{\textrm{km\,s}^{-1}}
\newcommand{\mpc}{\textrm{Mpc}}
\newcommand{\msun}{M_{\sun}}
\newcommand{\lsun}{L_{\sun}}

\newcommand{\hh}{\textrm{H}_{2}}

\begin{document}

\title{The Stellar Mass Content of Submillimeter-Selected Galaxies}

\author{Laura J. Hainline\altaffilmark{1,2}, A. W. Blain\altaffilmark{3,4},
Ian Smail\altaffilmark{5}, D. M. Alexander\altaffilmark{6},
L. Armus\altaffilmark{7}, S. C. Chapman\altaffilmark{8},
and R. J. Ivison\altaffilmark{9,10}}

\altaffiltext{1}{Dept.\ of Physics, United States Naval Academy,
572C Holloway Rd, Annapolis, MD 21402, USA; hainline@usna.edu}
\altaffiltext{2}{Dept.\ of Astronomy,
University of Maryland, College Park, MD 20742, USA}
\altaffiltext{3}{Dept.\ of Physics and Astronomy, University of Leicester, 
University Road, Leicester LE1 7RH, UK}
\altaffiltext{4}{Dept.\ of Astronomy, California Institute
of Technology, Mail Code 249-17, Pasadena, CA 91125, USA}
\altaffiltext{5}{Institute for Computational Cosmology, Durham University,
South Road, Durham DH1 3LE, UK}
\altaffiltext{6}{Dept.\ of Physics, Durham University, Durham DH1 3LE, UK}
\altaffiltext{7}{Spitzer Science Center, California Institute of Technology,
Mail Code 220-6, Pasadena, CA 91125, USA}
\altaffiltext{8}{Institute of Astronomy, University of Cambridge,
Madingley Road, Cambridge CB3 0HA, UK}
\altaffiltext{9}{Institute for Astronomy, University of Edinburgh, 
Blackford Hill, Edinburgh EH9 3HJ, UK}
\altaffiltext{10}{UK Astronomy Technology Centre, Royal Observatory, 
Blackford Hill, Edinburgh EH9 3HJ, UK}

\begin{abstract}

We present a new study of stellar mass in a sample of $\sim 70$ 
submillimeter-selected galaxies (SMGs) with accurate spectroscopic redshifts.
We fit combinations of stellar population synthesis models and power laws
to the galaxies' observed-frame optical through mid-IR spectral energy 
distributions to separate stellar emission from non-stellar near-IR continuum.
The availability of spectroscopic redshifts significantly enhances
our ability to determine unambiguously not only the mass and luminosity
of SMGs, but also the presence and contribution of non-stellar emission to their
spectral energy distributions. By separating the stellar emission from
the non-stellar near-IR continuum, we find that $\sim 50$\% of our sample
have non-stellar contributions of less than 10\% in rest-frame
$H$-band and $\sim 10$\% of our sample have non-stellar contributions
greater than 50\%.  We find that the $K$-band luminosity of the non-stellar 
continuum emission is correlated with hard X-ray luminosity, indicating
an AGN origin of the emission.  Upon subtracting this AGN-contributed 
continuum component from all of the galaxies in our sample, we determine a lower
median stellar mass for SMGs than previous studies, $\sim 7\times 10^{10}\,\msun$. 
We use constraints of
the starburst time-scale from molecular gas studies to estimate the
amount of fading our sample would undergo if they passively evolve
after the starburst terminates.  The results suggest that typical SMGs,
while among the most massive galaxies at $z\sim 2$, are likely
to produce descendants of similar mass and luminosity to 
$L^{\ast}$ galaxies in the local universe.

\end{abstract}


\section{INTRODUCTION}

Soon after their discovery, the population of high-$z$, 
far-infrared (far-IR) luminous galaxies revealed in deep submillimeter 
surveys \citep[e.g.,][]{smail97,hughes98,barger98,eales99}
were proposed to represent the formation of (the most) massive, metal-rich spheroids 
\citep{lilly99,smail02,swinbank06}. Submillimeter-selected galaxies (SMGs) 
have thus tantalized observers as a potential 
opportunity to witness the formation of massive galaxies directly.
The far-IR luminosities \citep[$> 10^{12}\,\lsun$; e.g.,][]{chapman05,kovacs06,magnelli10,chapman10}
of these dusty, gas-rich galaxies \citep[e.g.,][]{frayer98,grevesmg} 
imply star formation rates in excess of $10^{3}\,\msun\,\textrm{yr}^{-1}$,
sufficient to form the stellar mass of an $L^{\ast}$ galaxy in 
$\sim 100\,\textrm{Myr}$, potentially explaining the seemingly 
uniformly-old stellar populations of local ellipticals 
\citep[e.g.,][]{bower92,kauffmann03} and massive galaxy populations at $z\sim 1-3$ 
\citep[e.g.,][]{franx03,steidel03,glazebrook04,daddi04}.  
Moreover, the observed clustering of SMGs \citep{blain04b,weiss09} 
is similar to that of massive galaxies 
in the local universe, and suggests that SMGs 
reside within $\sim 10^{13}\,\msun$ dark halos. Yet,  
the formation of such massive galaxies on such rapid 
time-scales is difficult to account for using theoretical 
models of galaxy formation
\citep[e.g.,][]{guiderdoni98,devriendt00,baugh05,swinbank08}.  

Determining the masses of high-$z$ SMGs, especially the stellar 
and gas masses,  
and their relationship to other populations of high-$z$ galaxies 
is thus crucial to understanding the link between SMGs and massive
galaxies in the local Universe and places much-needed
constraints on the $z=0$ descendants of SMGs.
Prior to the advent of the \emph{Spitzer Space Telescope}, however,
constraints on the stellar mass  were difficult
to obtain for SMGs, especially for individual galaxies, due to
significant obscuration at optical and near-IR wavelengths 
\citep[rest-frame UV and optical; see][]{smail04,swinbank_spec,pope05}. 
The rest-frame near-infrared (near-IR), which \emph{Spitzer} gives us the power
to probe for high-$z$ galaxies, is less affected by reddening 
than optical-band data, and thus is beneficial when 
looking at SMGs which suffer from extinction in the
rest-frame UV and optical bands.  In addition, because the 
near-IR luminosity is much less dependent on past star formation 
history than, for example, the optical-band luminosity 
\citep[e.g.,][]{kauffmann98}, the rest-frame near-IR continuum emission 
permits the best determination of total stellar mass in both young
and old stars.  

Using data from \emph{Spitzer}-IRAC, \citet{borys05}, \citet{ljhthesis},
\citet{dye08}, and \citet{michalowski10} have all
calculated stellar masses for individual SMGs by
fitting the observed-frame optical through mid-IR spectral energy 
distributions (SEDs), obtaining typical stellar masses ranging from
$10^{10.8}-10^{11.8}\,\msun$; \citet{ljhthesis}, \citet{dye08}, 
and \citet{michalowski10} also find evidence of significant stellar mass assembly 
prior to the SMG phase.  However, none of the previous stellar mass studies addressed 
the contribution of near-IR continuum excess to the SEDs of SMGs,
noted by \citet{borys05} and \citet{ljhthesis} in numerous SMGs 
in their samples, which may represent very hot dust heated by AGN. 
Contamination of the stellar masses by AGN could especially be a problem
for \citet{borys05}, who used a small, X-ray-detected SMG sample
which was thus possibly biased towards SMGs with stronger AGN contributions
(or lower X-ray obscuration).
While X-ray observations and mid-IR spectra of SMGs indicate that 
the majority of the population are not dominated
by an active nucleus (AGN), many SMGs host AGN activity
\citep{alexander03,alexander05,pope08,karin09}, which may not be
negligible for the purpose of stellar mass studies.  

We have published a \emph{Spitzer}-IRAC and MIPS survey
of a sample of $\sim 70$ radio-detected SMGs with spectroscopic redshifts 
\citep[][hereafter Paper~I]{ljh09},
a factor of 5 larger than the sample considered by \citet{borys05}.
In this paper we combine our IRAC data set from Paper~I with optical and near-IR 
photometry from the literature to analyze the rest-frame UV to
near-IR SEDs of the sample to determine stellar masses for
these SMGs, while accounting for a contribution to the near-IR luminosity
from hot, possibly AGN-heated, dust, expanding the analysis of
\citet{ljhthesis}.  With our large sample size, spectroscopic redshifts\footnote{Note 
that using optical through near-IR SED fitting to obtain both photometric
redshifts and stellar mass for SMGs, as done in \citet{dye08}, is 
problematic in that both the redshift and mass are sensitive to the 
choice of spectral template used.}, and treatment of non-stellar emission, we 
can make estimates of SMG stellar mass which better represent 
the stellar component than those estimates currently available 
from \citet{borys05} and \citet{michalowski10}.  Using our new estimates, we 
can evaluate whether SMGs contain enough mass to represent the 
formation of the most massive galaxies and make predictions for SMG descendants.

The structure of the paper is as follows: we begin in \S\ref{sec:sed_data_sec},
describing our SMG sample and data.  In \S\ref{sec:sedmodel} we discuss the methods
we use to determine stellar masses for our sample.  In \S\ref{sec:mstar_sec} we present
our results from the fits of the stellar population models to the 
observed SEDs.  In \S\ref{sec:mstar_discussion} we conduct  
independent checks on the stellar masses from
dynamical masses obtained through observations of 
CO rotational transitions, with the caveat that the dynamical masses
are also subject to systematic uncertainties arising from assumptions that
the gas traces the total mass and the galaxies are in virial equilibrium, then 
compare the stellar masses of SMGs to those of
other high-$z$ galaxy populations, and conclude by discussing the 
evolution of SMGs to the present day.  Throughout our analysis and 
discussion, we assume $\Omega_{\textrm{M}}=0.27$, $\Omega_{\Lambda}=0.73$, 
and $H_{0}=71\,\kms~\mpc^{-1}$ \citep{hinshaw09}.

\section{OPTICAL, NEAR-IR, AND MID-IR DATA}\label{sec:sed_data_sec}

In Paper~I, we present \emph{Spitzer}-IRAC measurements
for the spectroscopic SMG sample from 
\citet[][C05 hereafter]{chapman05}.  Their final spectroscopic sample 
contains 73 SCUBA galaxies (median $S_{850}=5.7\pm 3.0$\,mJy), 
each of which was required to be detected in very deep 
($1\sigma$ noise $\sim 5-11\,\mu$Jy), high-resolution ($\sim 1\arcsec$) 
Very Large Array (VLA) maps at 1.4\,GHz
to obtain a position accurate enough for optical spectroscopy.
The redshift distribution of the C05 sample, when corrected for spectroscopic
incompleteness, is well-characterized by a Gaussian with
median redshift of $z = 2.2$ and $\sigma_{z}=0.8 $.  None of the $z>1$
SMGs in the final sample analyzed in C05 are suspected to be strongly lensed.
We refer the reader to C05 and Paper~I for further details on the sample
and \emph{Spitzer} observations. 
We will exclude from our analysis 2 SMGs from the IRAC-observed sub-sample
whose radio counterparts are in doubt (and which were not detected by 
\emph{Spitzer}): SMM\,J131231.07+424609.0 and SMM\,J221725.97+001238.9.
We also exclude from our analysis the galaxies 
which are suspected to be low-$z$ lenses of background submillimeter
sources in C05, as in C05 and Paper~I.  The final sample which we
analyze in this paper thus contains 67 radio-detected SMGs, and
includes the entire sample of 13 X-ray detected SMGs analyzed by \citet{borys05}.
We note that a bias towards emission-line galaxies is
introduced into our sample since the SMGs are required to be
spectroscopically-identified: the redshift of an emission-line galaxy is
more easily identified than that of a galaxy without emission lines 
when the continuum flux is faint and has a low signal-to-noise ratio.
While one obvious consequence of this bias is that our sample is incomplete
in the ``redshift desert'' between $z \sim 1.2$ and $z \sim 1.8$, the bias 
may also result in preferential selection of galaxies which contain 
less-obscured star formation or AGN activity.
In addition, since we require a radio detection of the galaxies in our sample
in order to obtain the spectroscopic redshift, we may be selecting against
SMGs with cold dust or higher redshift (see, e.g., Ivison et al.\ 2002;
Eales et al.\ 2003; C05).
As a result of these biases, the conclusions we draw must be 
restricted to apply only to spectroscopically identified, radio-detected SMGs.  However,
a wide range of luminosity and redshift is still probed within
these selection limits \citep[see, e.g.,][]{ivison02,blain04}.

Although we postpone a discussion of the uncertainties associated with
our modeling procedure to later sections, we note that, in principle, 
to obtain the best constraints on the stellar population ages and 
the extinction of starlight from fitting stellar population models to the
SEDs of high-$z$ galaxies, we must use photometry which spans the 4000\,\AA\
break and the rest-frame UV through optical bands, 
respectively \citep[e.g.,][]{shapley01,shapley05}.  
Thus, for our present study we combine observed-frame
optical and near-IR measurements from a variety of literature sources 
with our IRAC measurements presented in Paper I to obtain the most wavelength 
coverage possible.  The available observed-frame optical photometric bands vary for
each of the seven sky fields in our sample; see \S\ref{sec:sys_unc} for a discussion
of the impact on our final results of the varying number of photometric constraints within
the galaxy sample.   In Table~\ref{tab:smg_fields} we summarize 
for each field the photometric bands available and the typical number of 
optical and near-IR photometric constraints per galaxy.
For the CFRS-03h field, we use here $UBV$ data from \citet{clements04} and 
$R$ data from C05.  For the Lockman East field, we use the $B$ data 
when available from C05 and $R$ data from \citet{ivison05}.
In the HDF/GOODS-N field (hereafter simply the GOODS-N field), we
use the optical photometry catalogue of \citet{capak04}, which contains
measurements in $U$, $B$, $V$, $R$, $I$, and $z^{\prime}$.  The SMGs
in the SSA-13 field have $B$ and $R$ measurements in C05, plus 
$z$ band measurements from \citet{fomalont06}\ of comparable depth. In the CFRS-14h
and SSA-22 fields we use the $B$ and $R$ measurements from C05.  For the ELAIS-N2 field,
we take our optical data from \citet{ivison02}, in $B$, $V$, and $R$. 
The available near-IR data for our sample are more homogeneous, because we obtain 
all of the near-IR data ($I$, $J$, $K$) from the study of \citet{smail04}\
for all fields with the exception of the GOODS-N field. 
For SMGs in GOODS-N, we use the $I$ data of \citet{capak04}, but
the $J$ and $K$ data of \citet{smail04}.
We convert all the optical data obtained from the literature to total 
magnitudes using aperture corrections, generally supplied by the corresponding authors
of each data set, to eliminate the effects of differing seeing conditions
and photometry aperture size between different data sets.

We note that 4 out of the 67 SMGs from our IRAC-observed sub-sample of C05 SMGs 
have less than 3 detections across the observed-frame 
optical--mid-IR wavelength range.  While it is difficult to place meaningful 
constraints on their stellar SED to derive ages and extinction 
with so many upper limits, we can still
place upper limits on their luminosities in a given band, and use mass-to-light
ratios derived for other SMGs to place upper limits on their stellar mass.  Thus,
we retain these galaxies in our sample despite the small number of detections.

\section{CALCULATING THE STELLAR MASS OF SMGs}\label{sec:sedmodel}

Our approach to determining the stellar masses of SMGs is derived from
that of \citet{borys05}, because, as we discuss below in 
\S\ref{sec:fit_complications}, we find that the method typically
applied to obtain the stellar mass of intermediate- and high-redshift galaxies (i.e., fitting a 
stellar population model to each galaxy's SED to determine its 
stellar population age, visual extinction, and appropriate mass-to-light ratio) 
is unable to adequately constrain the stellar mass of individual
SMGs in our sample.  Rather than assume that the stellar population parameters 
and masses derived for individual SMGs from SED fitting are correct
in an absolute sense, we derive constraints on the stellar population 
parameters and stellar masses from averages over the SMG sample (\S\ref{sec:mass_est_proc}).  
In doing so, we can only constrain the stellar mass of
SMGs in an average sense (the mean and median of the sample).  
Although we will still present estimates of stellar mass for the individual 
galaxies for those who wish to use them despite the large uncertainties,
we will only discuss the average properties of the sample.

In the discussion that follows, and to obtain the average stellar
population parameters of SMGs, we have used version 
11.0 of the \textsc{hyper-z} photometric redshift software package 
\citep{hyperzref} to fit evolutionary population synthesis 
models to the optical--mid-IR SEDs of the individual galaxies in our sample.
\textsc{hyper-z} employs $\chi^{2}$ minimization to derive the best-fit
age/time since the start of star formation, optical extinction $A_{V}$,
and normalization for a given stellar population model.  We restricted
the code to fit the SEDs at the spectroscopic redshift of each SMG, 
and required that the best-fit ages be less than the age of the Universe 
at the redshift of each SMG.  We have fitted the population synthesis
models of both \citet[][using the Padova 1994 stellar tracks; hereafter BC03]{bc03} and 
\citet[][hereafter M05]{maraston05}, although in \S\ref{sec:mstar_sec} we
choose to report results for only one set of models.
We utilized the solar metallicity models based on the findings of \citet{swinbank_spec} that 
SMGs have metallicities consistent with those of UV-selected galaxies of similar
redshift, which for the most massive examples have approximately
solar metallicities \citep{erb06a}, and the findings of \citet{tecza04}.  
The BC03 and M05 models are made available with two different 
initial mass functions (IMFs):
the BC03 models have a choice of the \citet{salpeterimf} IMF or the
\citet{chabrier03} IMF, while for the M05 models we must choose 
between the \citet{salpeterimf} IMF and the \citet{kroupa01} IMF.
We preferred to not use the IMF of \citet{salpeterimf} since
observations of local galaxies suggest that this IMF over-predicts
the numbers of low-mass stars and thus the mass-to-light ratio
\citep[e.g.,][]{blain99,bell01}, casting suspicion on its use in high-$z$ galaxies.
Instead, we use the \citet{chabrier03} IMF for the BC03 models with a 
lower mass cutoff of $0.1\,\msun$ and upper mass cutoff of $100\,\msun$, and the 
\citet{kroupa01} IMF for the M05 models with the same low-mass cutoff.
Although the use of two different IMFs may be somewhat confusing,
the two different IMFs predict similar spectral properties
and mass-to-light ratios at a given age (BC03), and \citet{tacconi08}
find that the two IMFs produce similar results to within 
15\%.   To account for dust extinction in our SED fitting, we assumed a simple
dust screen model, although it is likely that SMGs contain a variety
of clouds of different obscuration attenuating the total observed starlight.  
We used the \citet{calzetti00}\ extinction law for starburst galaxies, 
allowing the extinction to vary within a range of $A_{V} = 0-5$\,mag.
In addition to extinction intrinsic to each galaxy, we also corrected
for the modest reddening along the line of sight to each SMG due to the Milky
Way using the dust maps of \citet{schlegel98}.  The line-of-sight
reddening estimate for each sky field is listed in Table~\ref{tab:smg_fields}.

\subsection{Complications In Applying Standard SED Modeling Techniques to Estimate
$M_{\star}$ for SMGs}\label{sec:fit_complications}

\subsubsection{Constraining Star Formation History of SMGs}\label{sec:sfhistory}

\begin{figure*}
\begin{center}
\includegraphics[scale=0.9]{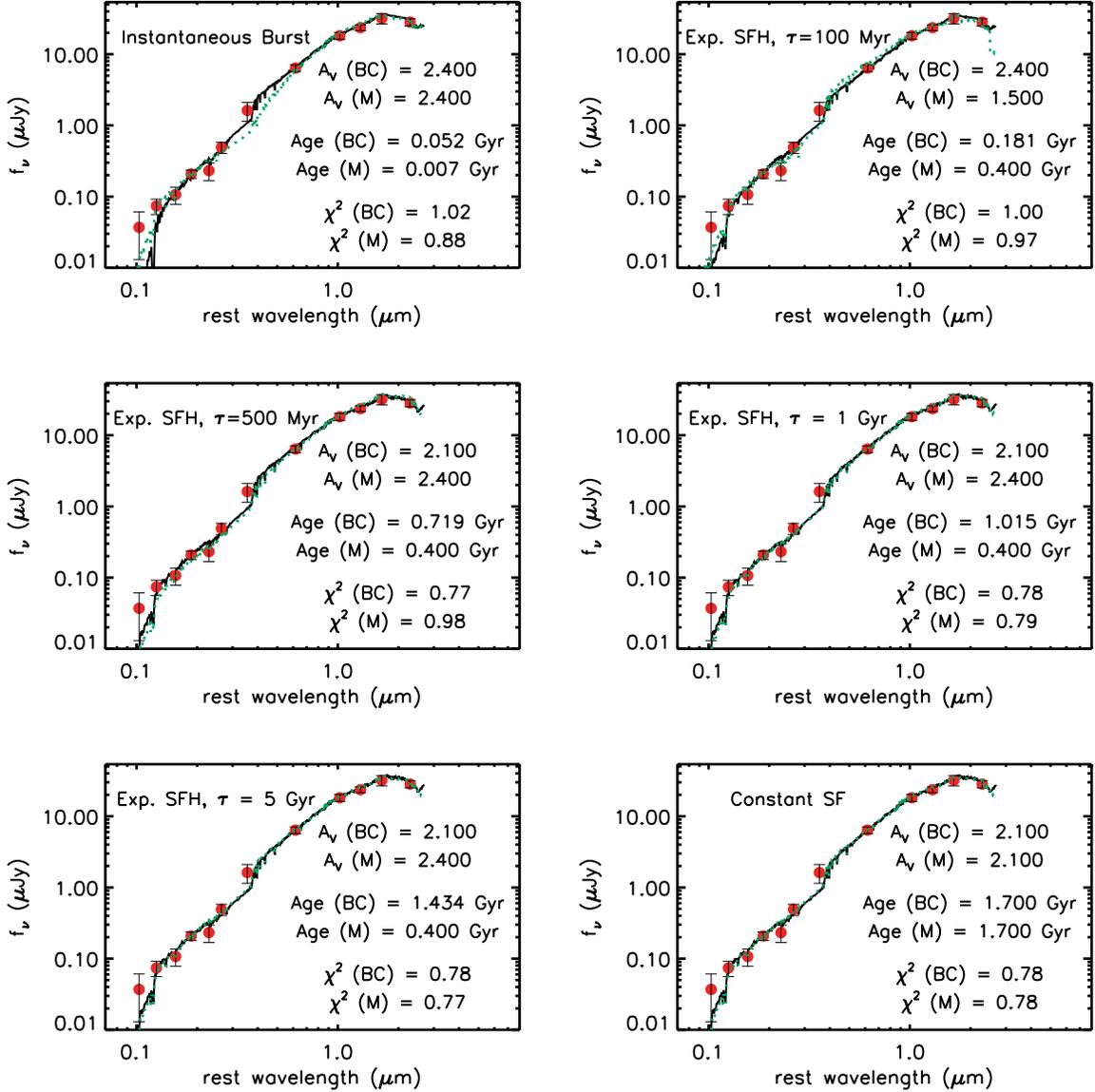}
\caption{
Fits of several different stellar population models of different star formation
history to the observed SED of SMM\,J123707.21+621408.1 (round points)
at $z=2.484$.  The
best-fit BC03 model spectrum is over-plotted as the solid black line, while
the best-fit M05 model is represented by the dotted (green/gray) 
line. The different fits are nearly equally acceptable in 
a $\chi^{2}$ sense.  Note that this particular SMG does not display a near-IR
continuum excess.}\label{fig:hizallmodels}
\end{center}
\end{figure*}

For our stellar mass study, we had originally intended to allow 
\textsc{hyper-z} to choose the best-fitting star formation history
for each SMG from a set comprised of an instantaneous 
starburst star formation history (which is equivalent to a simple stellar
population or SSP), a continuous star formation history,
and exponential star formation histories of the form
\begin{equation}
\textrm{SFR}(t) \propto \exp(-t/\tau),
\end{equation}
where $\tau = 0.1, 0.25, 0.5, 1, 2$, and 5\,Gyr, and thus obtain 
an unambiguous stellar population age, $A_{V}$, and stellar mass-to-light 
ratio which could be used to calculate the stellar mass of each galaxy. 
However, careful inspection of our tests utilizing the various star formation histories
reveals that for many SMGs in our sample, the fits could not
indicate a unique best-fit star formation history.
We find that the values of
$\chi_{\nu}^{2}$ corresponding to the best fit $A_{V}$, age, and 
SED normalization are extremely similar between even just 
this restricted set of star 
formation histories, as are the appearances of the fits. 
Admittedly, for some SMGs in the sample with relatively few
photometric constraints or anomalous observed-frame
optical photometry, such a result is not unexpected; however,
we encounter the same ambiguity among galaxies with the most
complete and accurate photometric data.
We illustrate the problem in Figure~\ref{fig:hizallmodels}, where we show
the observed SED of a typical SMG at $z\sim 2.5$ with excellent photometry, 
fit by a range of different simple star formation 
histories, using models of both BC03 and M05. 
Despite the broad range of photometric data in 
Figure~\ref{fig:hizallmodels} ($UBVRIz^{\prime}JK$+IRAC), nearly all of the
star formation histories can produce fits of similar likelihood 
(e.g., 4 of the 6 histories shown produce $\chi_{\nu}^{2}$ values within
$\pm 0.01$ of each other for the BC03 models, and all of the $\chi_{\nu}^{2}$ values
correspond to a probability of 40--70\% of exceeding $\chi^{2}$ for
the number of degrees of freedom), 
with ages varying from $\sim 10$\,Myr to $\sim 2$\,Gyr and extinction
in the range $A_{V}=1.5-2.4$\,mag.  

Thus, we conclude that we are unable to place useful constraints 
on the star formation histories of individual SMGs through fitting
such simple stellar population models, as \citet{shapley01,shapley05}, \citet{papovich01},
and \citet{erb06} also find for UV-selected high-redshift galaxies 
with similarly broad wavelength coverage.  Since the best-fit ages 
are strongly dependent on the assumed star-formation 
history \citep[the instantaneous burst models tend to produce the youngest ages, 
while the continuous star formation models tend to produce the 
oldest ages, see, e.g.,][]{maraston10}, we are also 
unable to effectively constrain the ages of individual galaxies and thus
the appropriate mass-to-light ratio of the stellar populations in 
individual SMGs without other information on the likely star formation
histories, introducing extra systematic uncertainties into 
the stellar mass calculations,
on top of any resulting from the stellar models themselves.

Some of our inability to constrain the star formation history and the stellar
population ages stems from the available photometric data for SMGs.
A small subset of SMGs in our sample (10 out of 67) possess 
seemingly anomalous observed-frame optical photometric data; 
for these galaxies, the universally-high $\chi_{\nu}^{2}$ values indicate that
none of the stellar population models adequately describe the SED\footnote{The causes
of the anomalous photometry are unknown, as no notes are available from the literature
data sources for these galaxies.  In a few cases, the rest-frame wavelengths of the discrepant
data suggest a contribution from an emission line; however, in the remaining cases
the optical counterpart could have been mis-identified or the photometric error is 
larger than stated.}.  Yet, even when the wavelength coverage and data quality 
are excellent (e.g., the GOODS-N field), we still may have insufficient information. 
The determination of age from stellar population models relies heavily 
on the shape of the Balmer break or 4000\,\AA\ break in the SED 
\citep[see, e.g.][]{kriek08,whitaker10}\footnote{\citet{kriek08} show, using their
sample of $z\sim 2$ massive red galaxies with spectroscopic redshifts, that
the best-fit ages obtained through stellar population synthesis modeling of
SEDs constructed using broadband $JHK$ photometry have a mean absolute deviation of a
factor of $\sim 3$ relative to the ages derived from SEDs constructed using low-resolution
near-IR spectra.}.   Unfortunately, for only up to a third of SMGs in our 
sample do we see a suggestion of a break in the continuum near 4000\,\AA, and we 
usually do not have enough photometric data points for our SMGs in the 
critical area of the Balmer/4000\,\AA\ break to clearly delineate the location
and the shape of the spectral break.  
In addition, the well-known degeneracy between age 
and extinction \citep[e.g.,][]{sawicki98,shapley01}, in 
which a reddened, young starburst population can look similar to an
older, unextincted stellar population, is also probably contributing to
the ambiguity in the stellar ages.  However, we expect that much of the 
difficulty in constraining stellar population ages is likely to arise 
from the application of relatively simple stellar models to what is likely
an intrinsically complex mix of stellar populations of differing 
metallicity, reddening, and age.  Accordingly, we could make use of
stellar population models which employ more complex star formation histories,
such as short bursts on top of an exponential star formation history 
\citep[e.g.,][]{borch06,kaviraj07,dye08b};
however, we feel that doing so is not justified because the observational 
data are insufficient to constrain even simple star formation 
histories and cannot provide a unique solution.  We also stress that 
the uncertainties here, for SMGs with known redshifts, are far less
than if the redshift also has to be derived from the photometric fit.

\begin{figure}
\includegraphics[width=1.0\columnwidth,trim=40pt 10pt 10pt 20pt,clip]{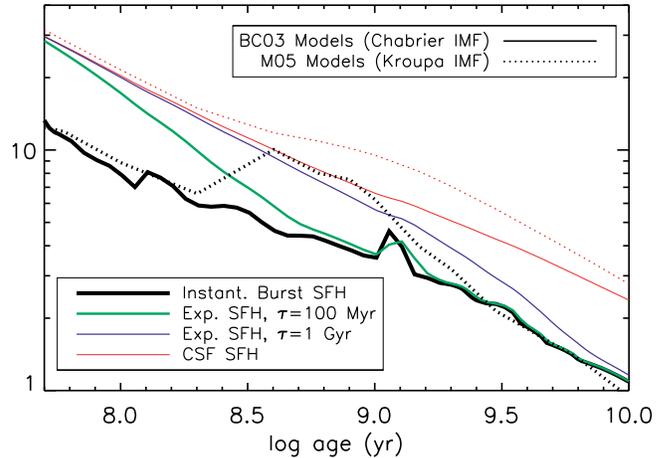}
\caption{$L_{H}/M_{\star}$ as a function of time since onset of star formation
for solar-metallicity population synthesis models of BC03 and M05.
Solid lines represent models of BC03, while dotted lines represent
the models of \citet{maraston05}.  The increase in $L_{H}/M_{\star}$ near
$\log{\textrm{age}} = 8.5$ in the \citet{maraston05}\ burst model marks the onset
of the TP-AGB phase.  Because at any age in between 50\,Myr and 3\,Gyr the 
$L_{H}/M_{\star}$ of the star formation history with typically the lowest $L_{H}/M_{\star}$ ratio (IB) differs 
from that of the star formation history with the highest typical
$L_{H}/M_{\star}$ (CSF) by at most a factor of 2--3, we can constrain the stellar masses
of individual SMGs to within a factor of 2--3 without knowing the star 
formation history.}\label{fig:lmratio_vs_time}
\end{figure}

The star formation history, and by extension, age, of a stellar population are critical
elements in determining the appropriate $M/L$ ratio needed to 
accurately convert a single galaxy's stellar luminosity to stellar mass.  However,
if one is willing to forego accurate masses for individual galaxies,
one may take advantage of statistics to obtain reasonably 
precise average and typical stellar masses from a large sample of 
individual galaxy masses which are much more uncertain. 
Fortunately, the light-to-mass ($L/M$) 
ratios of stellar populations of different star formation histories at ages above 50\,Myr,
shown in Figure~\ref{fig:lmratio_vs_time} for a sample of star formation histories,
have a narrower range of values than the ages themselves and are thus less uncertain
than the ages: for example, in the M05 stellar population models,  
the $L/M$ ratio of an instantaneous burst (IB hereafter) population
with age 100\,Myr differs by only 20\% from that of a 
constant star formation (CSF hereafter) 
population with age 2\,Gyr.  At any age in the range 50\,Myr to
3\,Gyr (approximately the age of the universe at $z\sim 2$), 
the $L/M$ of the star formation history with typically the lowest $L/M$ ratio
(IB) differs by only a factor of 2--3 at most
from that of the star formation history with the highest typical
$L/M$ (CSF).  So, with no constraint
on the star formation history, we can still constrain the stellar masses
of individual galaxies to within a factor of 2--3, neglecting other systematic
uncertainties.   If we assume the star formation histories of individual SMGs are 
uniformly distributed over the range of possible histories, which is 
reasonable because our attempts to fit the SEDs
of SMGs with a variety of star formation histories revealed no single preferred 
history over the entire sample, the average star formation history for the
population and the average $L/M$ ratio should exhibit less
dispersion than those associated with individual galaxies.   Consequently, the 
uncertainty in the sample-averaged stellar mass derived using the average $L/M$ ratio 
will be much less than the factor of 2--3 uncertainty in the individual masses.

In the absence of any indication of which star formation history is most
appropriate, and with the knowledge that (a) the SED fits of the 
different star formation histories to our SMG sample are correlated; 
(b) the best-fit ages are likely to not be representative of the true
stellar population since the light we receive is strongly modified by reddening;
and (c) the $L/M$ ratios implied by the different star 
formation histories will be similar to within a factor of a few,
we elect to compute stellar masses for our sample which are
averages over different possible star formation histories, instead
of basing our computation on a particular star formation history.  We
adopt the approach of \citet{borys05}, in which we assume a 
single $L/M$ ratio for the entire SMG sample for 
each of the IB and CSF histories, and average the resulting masses.  We choose these two because 
they are expected to represent the lower and upper extremes of the stellar
properties predicted by the different star formation histories;  
in averaging their results, we hope to reduce the 
systematic effects of star formation history so that our stellar 
mass estimates are correct to within a factor of 2.  The mean and median of
the stellar masses calculated in this way of the full SMG sample
will thus be the best-constrained quantity we obtain.

\subsubsection{Near-IR Excess in SMGs and Its Removal}\label{sec:excess_desc}

In Paper~I, we found that numerous galaxies in our SMG sample 
display evidence of non-stellar emission in their rest-frame
$J-H$ and $H-K$ colors (see Figure~6 in Paper~I), by comparing to samples
of ``normal'', non-active local galaxies and local ULIRGs.  Upon close 
examination of the observed SEDs of our sample, we find that
approximately 20\% of the sample display a clear red upturn in
the rest-frame near-IR portion of their SED.  The red upturn
suggests the presence of hot dust and possibly an obscured AGN, 
since stellar SEDs typically fall past the $1.6\,\micron$ stellar peak.  
An additional 50\% of the SMGs in the sample display 
smaller near-IR excesses redward of the $1.6\,\micron$ stellar peak
similar to those found in 24\,$\micron$-bright luminous IR galaxies and $K$-band selected galaxies
at $0.5 < z < 2.0$ by \citet{magnelli08} and \citet{mentuch09}.  These authors
suggest that the excess arises from very small, hot dust grains in star formation
regions, comparable to the near-IR emission detected from Galactic cirrus
\citep{flagey06} and reflection nebulae \citep[e.g.,][]{sellgren83} as well as local star-forming
galaxies \citep[e.g.,][]{lu03}, which can be described by graybody emission
at temperatures $\sim 1000$\,K.  
56\% of our sample also meet the criterion of \citet{alonsoherrero06} 
for IRAC power-law galaxies (PLGs; power law slope $\alpha > 0.5$), which those authors consider to
be obscured AGNs.  Yet, it is not completely clear that the PLG classification
can generally be attributed to AGN since \citet{donley08} find that star forming 
ULIRG templates at $z>2.6$ meet the PLG criteria as well (indeed, three of the four
SMGs that show no evidence of SED upturn but have $\alpha >  0.5$ are at
$z > 2.8$).  Regardless of whether or not the origin of the observed 
excess continuum emission is an obscured AGN or star formation regions, its contribution
to the near-IR emission of $\sim 70$\% of the SMGs in our sample must
be removed in order to derive the most representative stellar
luminosity and mass.  

We remove the non-stellar continuum contributions to the near-IR 
emission of our SMG sample by assuming that the total observed SED
can be represented as a sum of a stellar population model and 
a power law (PL) of the form $f_{\nu} \propto \nu^{-\alpha}$ representing the excess,
where $f_{\nu}$ is the flux density at frequency $\nu$. 
Note that, if the excess is in fact due to graybody dust emission in star forming regions
rather than dust heated by an obscured AGN, the PL model will still
be approximately correct, since the modified Planck function 
describing a graybody can be approximated as a PL over a small 
range of wavelength (e.g., a graybody of temperature 700--1000\,K 
and emissivity $\epsilon_{\nu} \propto \nu^{1}$, appropriate for small
dust grains in star-forming regions, can be approximated
by a PL with index ranging from -1.8 to -4 in the wavelength 
range 2--3\,$\micron$, depending on the temperature).
However, if the excess is due to a 3.3\,$\micron$ polycyclic
aromatic hydrocarbon (PAH) emission
line, the PL model will not improve the fits.  We scale the PL to the IRAC 8.0\,$\micron$
data point, and denote this scaled PL as the maximum possible
PL contribution. We then subtract from \emph{all} the 
observed photometry increasing fractions 
of the flux in this scaled PL, in steps of $0.1\times$ (maximum PL
contribution) and fit the BC03 and M05 stellar population 
models to the PL-subtracted data points.  The PL 
fraction which results in the lowest value of $\chi_{\nu}^{2}$
gives us the luminosity of stars only, which we can then use to find the stellar
mass.  We estimate the error in the PL fraction by determining
the PL fractions which result in a change of $\Delta \chi_{\nu}^{2} =  1$
above and below the best-fit PL fraction.

The choice of the best PL index to use in our PL removal
is not obvious, because differing obscuration levels can cause the 
rest-frame near-IR PL slope of hot dust emission 
to vary between galaxies even when the AGN powers
the dust emission \citep[e.g.,][]{simpson00}.
Observationally-derived near-IR PL slopes for local unobscured,
type-1 AGN have a typical value of $\alpha \sim 1 - 1.5$ 
\citep[e.g.,][]{malkan83,edelson86,neugebauer87}, whereas the range
for local Seyfert\,2 nuclei is $\alpha \sim 2 - 4$ \citep[e.g.,][]{alonsoherrero03}.
Obscured starbursts will have similar near- and mid-IR PL slopes to
obscured (type 2) AGN:  as mentioned in the previous paragraph,
graybody dust emission from star forming regions can be approximated
over $\lambda \sim 2-3\,\micron$ as a PL with $\alpha \sim 2 - 4$. 
We initially chose to use a PL index of $\alpha = 2$, based on
the observed average mid-IR continuum slope for SMGs with IRS spectra, 
$\alpha_{\textrm{MIR}} = 1.97\pm 0.22$ \citep{karin09}. 
However, for some galaxies in our sample, the $\nu^{-2}$ PL
is too shallow and results in a very poor fit; thus, 
we have allowed the PL in our continuum removal to have the form
$\nu^{-2}$ or $\nu^{-3}$, whichever best fits the data 
in a $\chi^{2}$ sense for each SMG.  By allowing 2 different possible
values of the PL index, we introduce a additional uncertainty
into our stellar masses; however, the uncertainty in the average mass 
is small (see \S\ref{sec:sys_unc}).  The results of our SED fits 
suggest that $\alpha = 3$ typically produces a better fit to the 
observed data overall: for the 82\% (72\%) of the SMG sample which 
have a non-zero best-fit PL contribution at observed-frame 8.0$\micron$
using the M05 IB (CSF) models, 70\% (79\%) have best-fit PL indices
of $\alpha = 3$ while 30\% (21\%) have best-fit PL indices of 
$\alpha = 2$.  When the BC03 stellar models are used in the SED fits,
the relative fractions change somewhat: for the 70\% (79\%) of the sample
with a non-zero best-fit PL fraction for the IB (CSF) star formation
history, 74\% (81\%) have best-fit PL indices
of $\alpha = 3$ while 26\% (19\%) have best-fit PL indices of 
$\alpha = 2$.

We note that when we 
extrapolate the observed-frame 24\,$\micron$ flux from the 8.0\,$\micron$ flux
based on the best-fit PL fraction, the extrapolated value exceeds the
observed 24\,$\micron$ flux in $\sim 50$\% of cases when $\alpha=3$ 
(the median over-prediction for that 50\% of the sample is a factor of 2.3)
and for $\sim 20$\% of the sample when
$\alpha=2$ (by a median factor of 1.5).  The $\sim 50$\% and 
$\sim 80$\% which are not over-predicted by $\alpha=3$ and $\alpha=2$ PLs,
respectively, are under-predicted. The over-prediction
of the rest-frame mid-IR flux based on the rest-frame near-IR PL
implies that the near-IR and the mid-IR require different PL slopes.  
This phenomenon is observed in local Seyfert galaxies 
\citep[e.g.,][]{edelson87,alonsoherrero03}, 
and can occur when an unobscured nucleus' emission is modified by a foreground reddening
screen (e.g., its host galaxy).  Such curvature in the mid-IR SED of
high-$z$ QSOs is also found by \citet{gallagher07}, who
find that the rest-frame 8.0\,$\micron$ luminosities of their sample are
over-predicted by factors ranging from 1.25--2.5 upon 
extrapolating the best-fit observed-frame 1.8--8.0\,$\micron$ 
PL to rest-frame 8.0\,$\micron$.  Thus, our choice of near-IR PL indices
for fitting are not necessarily invalidated by over-predicting the observed
24\,$\micron$ flux.

An alternate approach to fitting the near-IR excess emission
in the observed SED for the purpose of estimating stellar masses 
is to fit a combination of stellar population synthesis 
models and AGN (type 1 and/or 2) templates from the literature
\citep[see, e.g.,][]{merloni10}.   
However, we prefer the stellar population plus
PL approach over the use of templates for several important reasons.
First, because we have no spatial information for the near-IR excess
observed in our SMGs, and because SMGs are strongly star-forming
galaxies, we are hesitant to assume a priori that the excess arises from an
AGN when dust heated by star formation could also explain the observations.
The vast majority of our SMG sample are adequately 
fit in the rest-frame UV and optical by stellar models only, 
so it is not unreasonable to consider the case in which 
stellar radiation powers the near-IR excess.   Setting aside our
wish to not make assumptions regarding the origin of the near-IR excess,
our observation that the SMG SEDs can generally be described by 
stellar emission argues against a need for type-1 AGN templates; 
a case could still be made, though, for the use of type-2 AGN templates.
On the other hand, our stellar population
model plus PL representation is a good reproduction
of a type-2 AGN SED, since a type-2 AGN SED is dominated
by stellar light in the rest-frame optical and PL continuum
in the near-IR.  Moreover, by using our stellar model plus PL approach
it is actually more straightforward to separate the contributions of stellar
emission and featureless continuum to an SED than would be 
possible using a stellar model and a type-2 AGN template, 
because the type-2 AGN template may contain some contribution
from the AGN host galaxy in the near-IR.
Finally, and most importantly, we prefer to use our PL removal model rather than
AGN templates because the PL model permits some flexibility 
in the near-IR PL index.  We discussed above how our efforts to 
determine the most appropriate near-IR PL index for the removal
of the continuum excess revealed that we could not fit all of
the SMGs in our sample with a single PL index; PL index variation
in the near-IR is also observed in local AGN.  Yet, by using an
AGN template to fit the observed SEDs, we would have only one
option for the near-IR PL index, which may or may not be 
appropriate for our sample. For example, the average optically-bright
(type-1) QSO SED constructed from Sloan Digital Sky Survey QSOs by 
\citet{richards06}, frequently used as an AGN template, has a PL 
index $\alpha = 1.8$ in the wavelength range 1--3\,$\micron$.  This PL
index will somewhat under-subtract the near-IR continuum component in those SMGs which
are best fit by $\alpha = 2$ and significantly under-subtract
the continuum for the SMGs which are best-fit by $\alpha = 3$ 
(the majority of the sample).  The composite type-2 AGN
SED of \citet{polletta07}, another commonly-used AGN template, has a near-IR PL
index $\alpha = 1.3$ between 1 and 3\,$\micron$; this template will 
thus significantly under-subtract the continuum component in nearly 
our entire SMG sample. 

\begin{figure}
\includegraphics[width=1.0\columnwidth,scale=0.9,trim=0pt 5pt 0pt 10pt,clip]{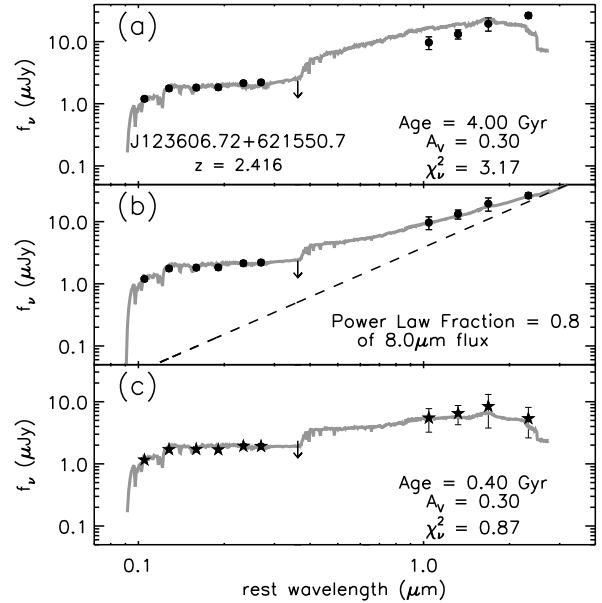}
\caption{
Example of stellar population model plus power-law fraction SED
fitting used for our sample of SMGs.  Panel (a):  The observed
$UBVRIz^{\prime}JK$+IRAC photometry (round points) for a typical
SMG with a near-IR excess is shown with the best-fit
CSF stellar population model of
M05 to that data.  The near-IR excess is most clearly seen
at $\lambda_{rest} > 2\,\micron$.  We note that the
requirement that the age of the best-fit model be physical was relaxed
in this particular case to allow the fit to converge.  Panel (b): The observed 
photometry for the same SMG is shown along with the best-fit
combination of $\nu^{-2}$ power law and CSF 
stellar population model from M05.  Panel (c):
The power-law subtracted photometry (black stars), representing the
total stellar contribution to the observed SED is shown with 
the best-fit CSF stellar population model
from M05 to those data points only.  Note from 
panel (a) that the unphysical age of the best-fit stellar 
population model (it is greater than the age of the universe at
$z=2.4$) and the statistically poor fit are indications 
that the stellar population model alone is insufficient to account
for the near-IR emission from this SMG.}\label{fig:agn_removal_ex}
\end{figure}

\begin{figure*}
\begin{center}
\includegraphics[scale=0.8]{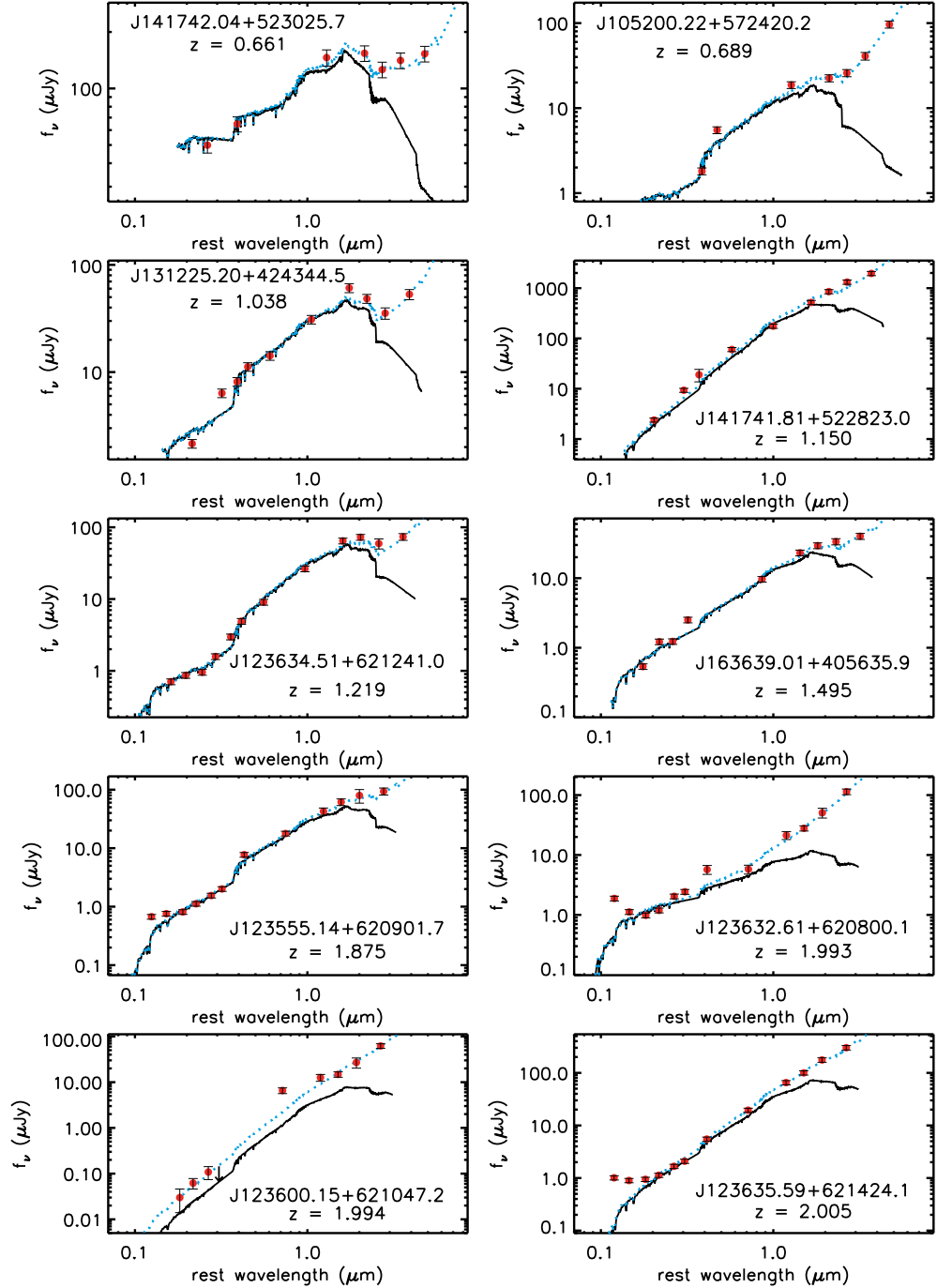}
\caption{
Rest-frame UV through near-IR SEDs of spectroscopically-identified SMGs
in our sample unambiguously requiring significant 
near-IR power law components to fit
the observed SED.  Round points are observed photometric data, 
while non-detections are indicated by downward arrows originating
from the measured upper limit.  The best-fit CSF model of
M05 is over-plotted as the black line.  The best-fit
combined stellar population model plus best-fit power law is
represented by the dotted (blue/gray) line.  The UV excess displayed by several of
these SMGs cannot be reproduced by stellar or near-IR power-law components
or a combination of the two.}\label{fig:iracseds1}
\end{center}
\end{figure*}

We illustrate the effects of our PL removal procedure on
a typical SMG in our sample showing a near-IR upturn in 
Figure~\ref{fig:agn_removal_ex}. 
We first show the observed optical--IRAC
photometry and the best-fit CSF 
stellar population model from M05 to the observed data; the 
$\chi_{\nu}^{2}$ of the fit indicates that the fit is poor.  We then show
the best-fit combination of $\nu^{-2}$ PL and CSF stellar
population model to the same observed photometry as well as the flux contributed
by the PL component.  Finally, we show
only the stellar component of the fit from the
middle panel which allows us to determine the stellar luminosity of
the SMG: the PL-subtracted photometry is plotted along with
its corresponding best-fit CSF model.  Note the significant improvement
in the $\chi_{\nu}^{2}$ statistic of the stellar model fit
when the PL component is subtracted as well as the more appropriate
age of the best-fit model.  For $\sim 60$\% of the SMGs
the best $\chi_{\nu}^{2}$ value of the fit improves when we subtract a 
PL contribution from the observed SED; indeed, for $\sim 20$\%
of the sample, the best $\chi_{\nu}^{2}$ of the fit decreases
by a factor of 2 or more when we subtract the PL contribution.
In Figure~\ref{fig:iracseds1} we show the results of the best CSF stellar population plus
PL fit to the observed photometry for the SMGs in our sample 
requiring PL contributions at observed-frame 8.0\,$\micron$ greater than 70\%,
for which the fits improved the most with the inclusion of the PL component.
In Figure~\ref{fig:rest_seds} we show the best CSF stellar model plus PL
fits for the remainder of the SMGs, for which the necessity of a PL
SED component can be more ambiguous.
We note that, for both Figures~\ref{fig:iracseds1} and \ref{fig:rest_seds},
the IB fits are visually similar to the CSF fits, although the best-fit PL 
contributions differ between the IB and CSF star formation histories for $\sim 60$\% of 
our sample for both the BC03 and M05 models.  In contrast to the similar
visual appearance of the IB and CSF fits, the best $\chi_{\nu}^{2}$ values for
the IB star formation history are 10\% greater (BC03 models) and
16\% greater (M05 models), on average, than
the best $\chi_{\nu}^{2}$ values for the CSF star formation history.

\begin{figure*}
\begin{center}
\includegraphics[scale=0.8]{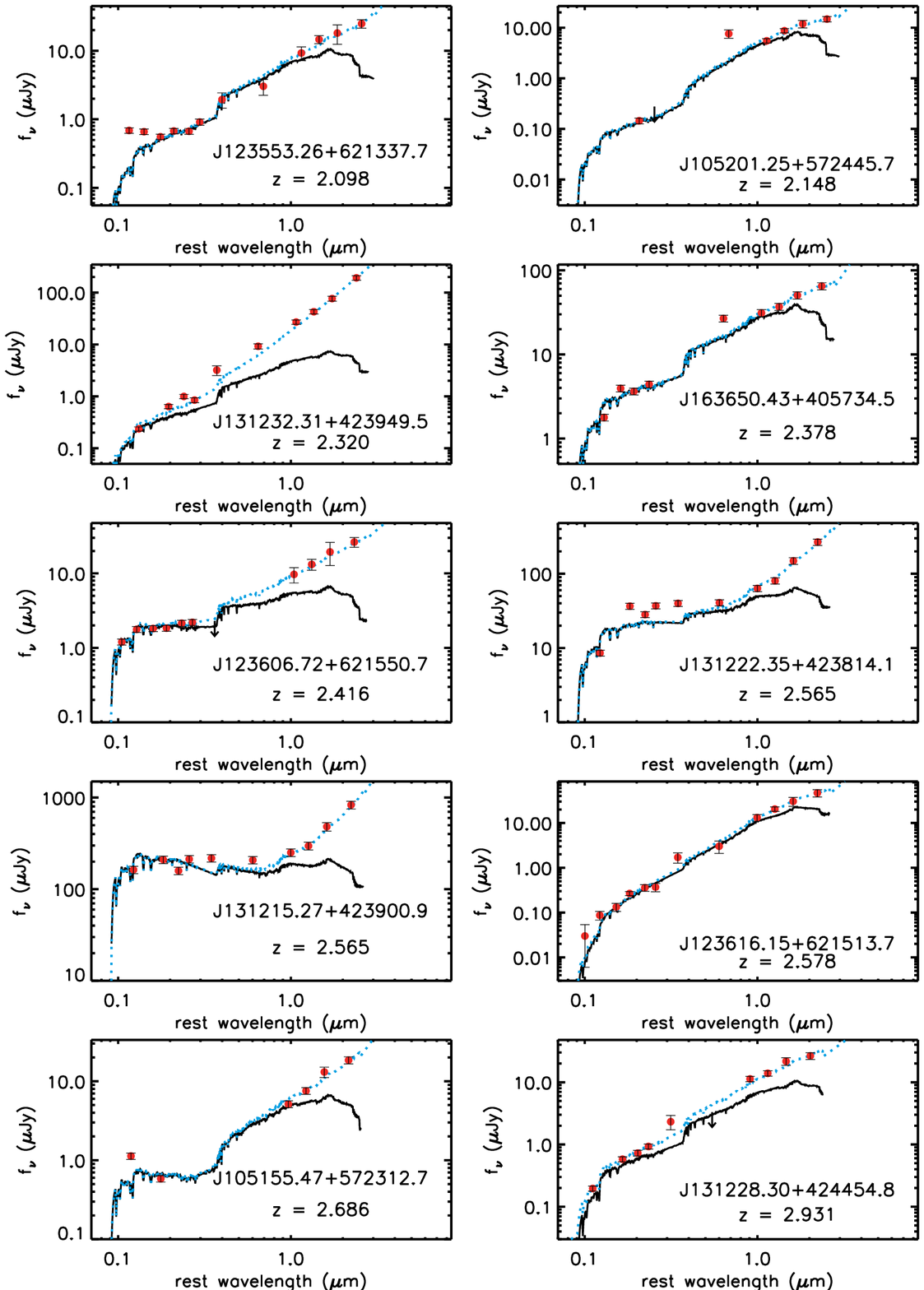}
\caption{Figure~\ref{fig:iracseds1} continued.}
\end{center}
\end{figure*}

\begin{figure*}
\begin{center}
\includegraphics[scale=0.7]{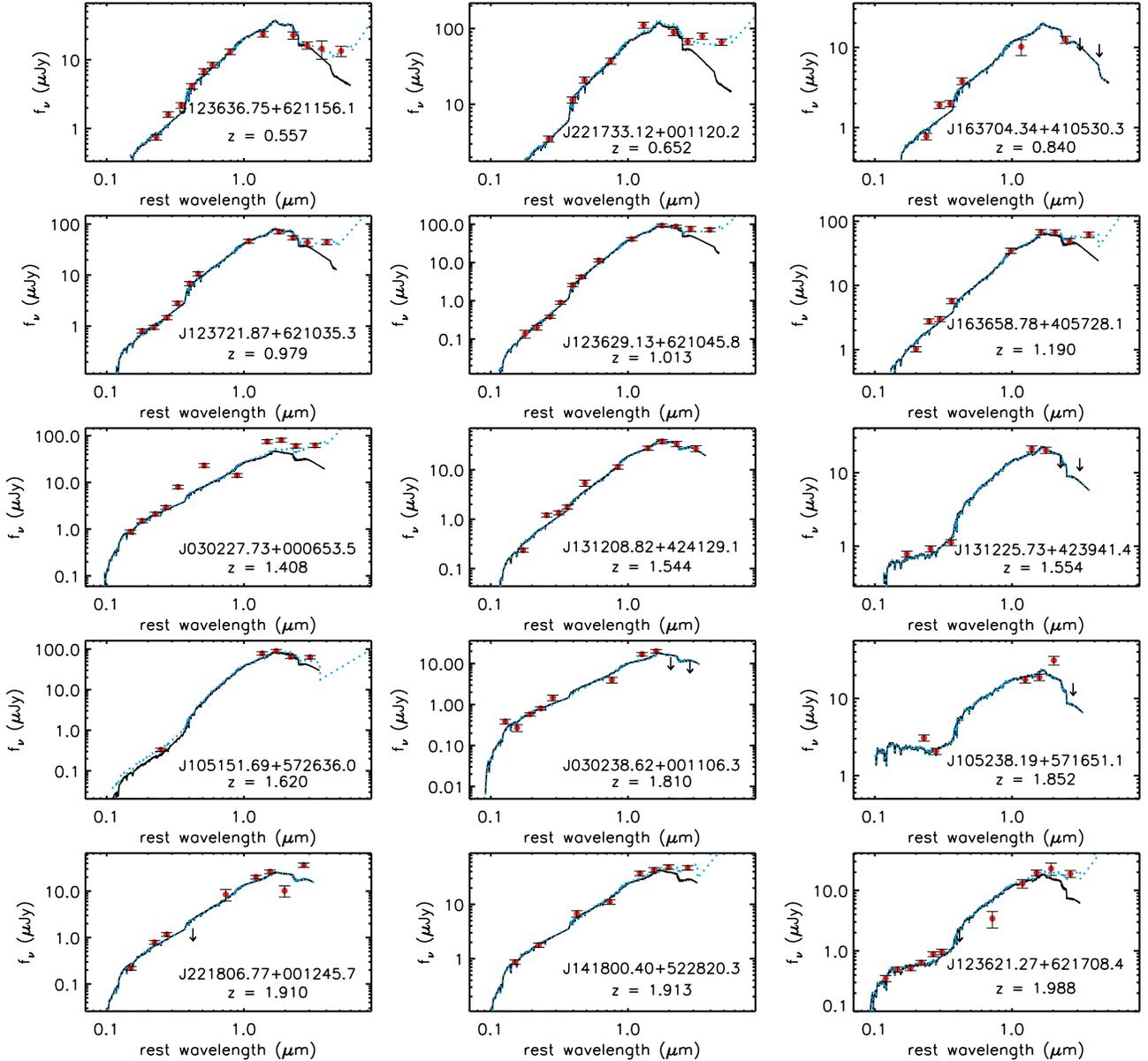}
\caption{Rest-frame UV through near-IR SEDs of 
remaining SMGs in our sample.
Symbols and lines are as in Figure~\ref{fig:iracseds1}.  
The large variation in non-stellar near-IR continuum contribution within the sample is 
apparent when comparing to Figure~\ref{fig:iracseds1}, ranging
from pure stellar emission to nearly pure power-law emission. }\label{fig:rest_seds}
\end{center}
\end{figure*}

\begin{figure*}
\begin{center}
\includegraphics[scale=0.7]{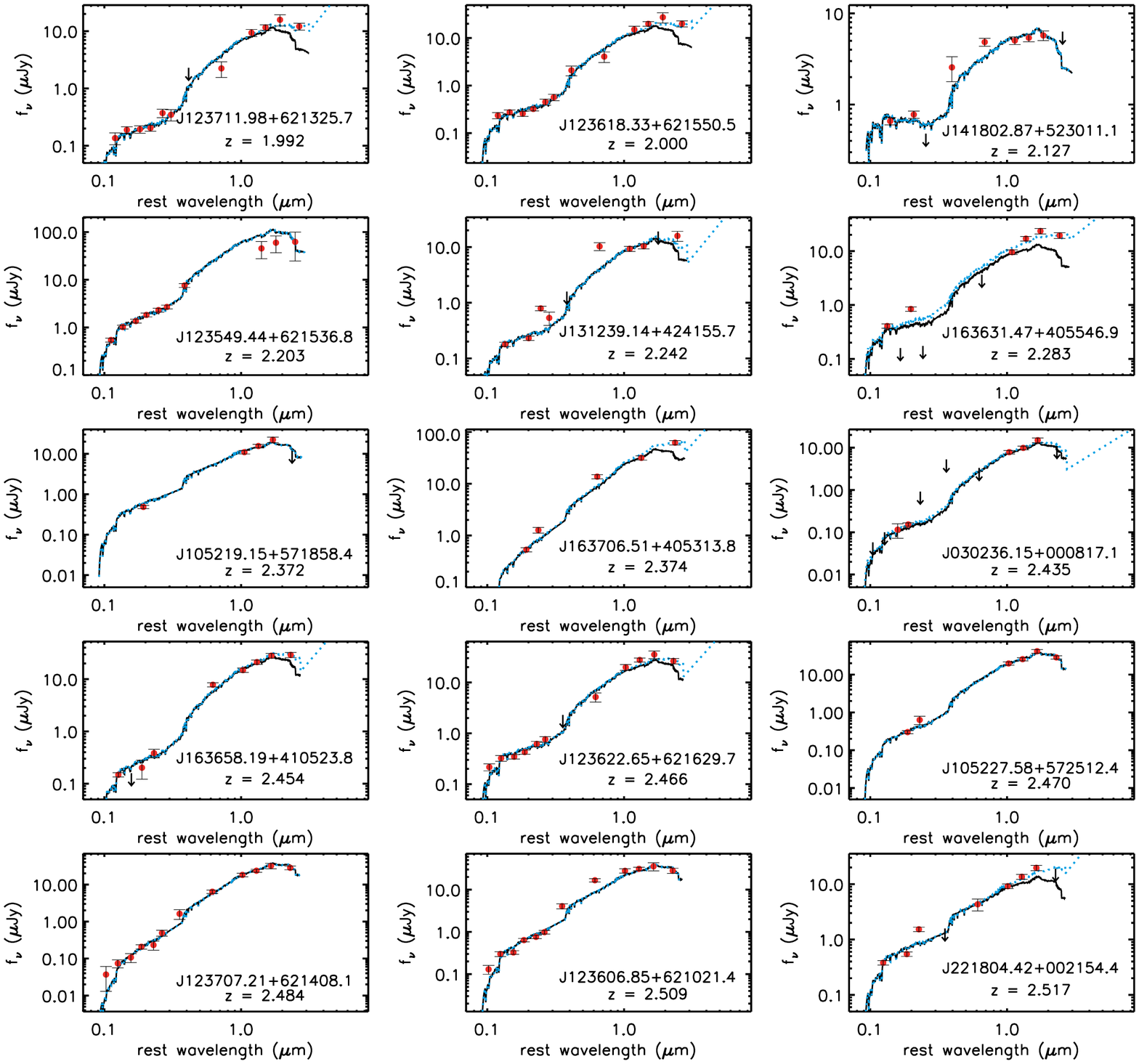}
\caption{Figure~\ref{fig:rest_seds} continued.}
\end{center}
\end{figure*}

\begin{figure*}
\begin{center}
\includegraphics[scale=0.7]{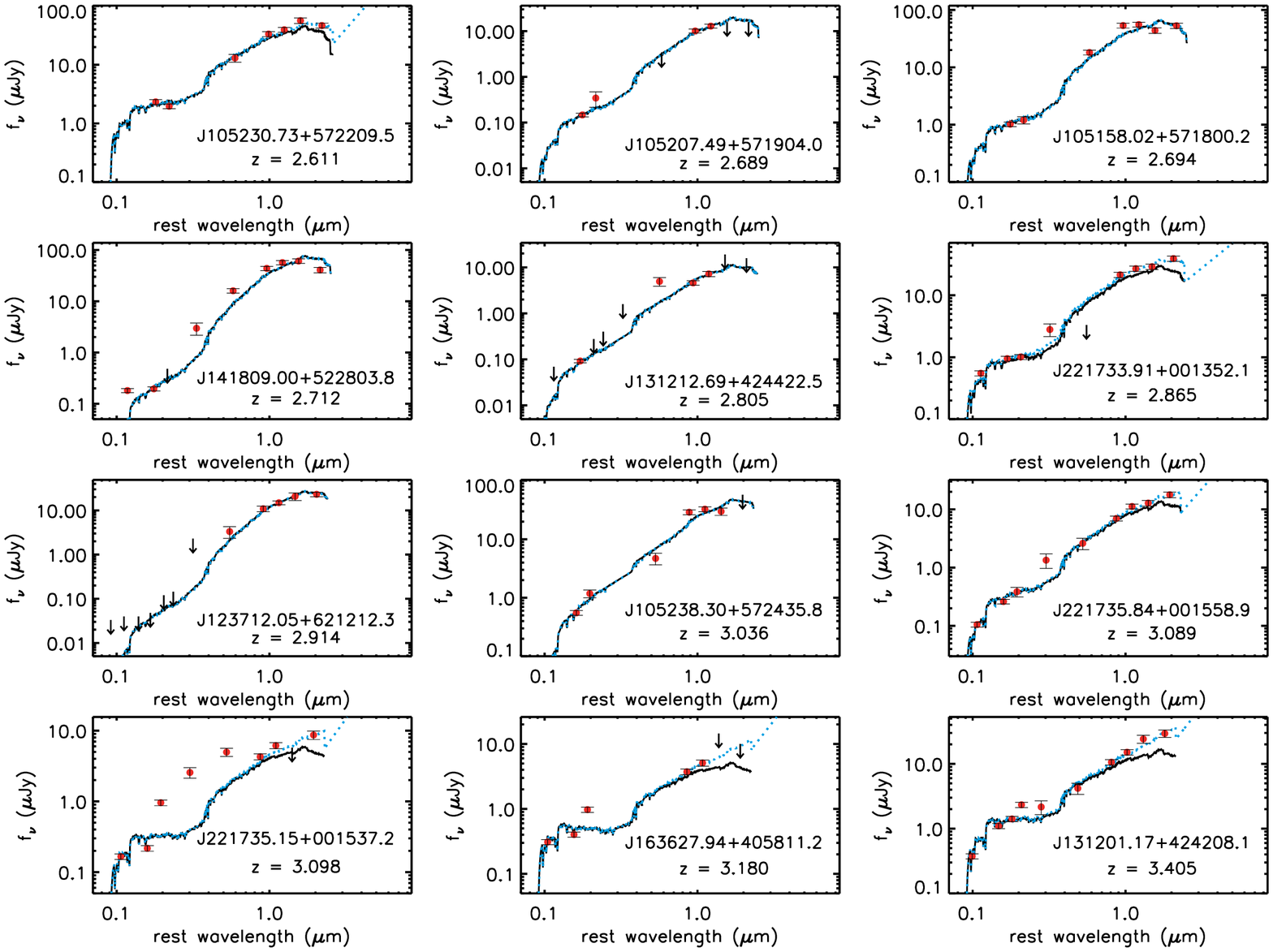}
\caption{Figure~\ref{fig:rest_seds} continued.}
\end{center}
\end{figure*}

\subsection{Stellar Mass Calculation Details}\label{sec:mass_est_proc}

We calculate a stellar mass ($M_{\star}$) for each SMG 
from the de-reddened, best-fit-power-law-subtracted absolute $H$ magnitude ($M_{H}$).  
$K$-band light has been used to determine the stellar mass
in past studies of the stellar mass of SMGs \citep[e.g.,][]{borys05,michalowski10}
because of its low sensitivity to previous 
star formation history and dust obscuration. We instead choose to use the 
rest-frame $H$ band to estimate $M_{\star}$ for our sample of SMGs 
to benefit from the low extinction of the near-IR bands while minimizing the
uncertain influence of thermally pulsating asymptotic giant branch (TP-AGB) 
stars on the model $K$-band light-to-mass ($L/M$) ratio in 
the M05 stellar population models.  
The change in $H$-band $L/M$ ratio induced by the appearance of a dominant 
solar-metallicity TP-AGB population at ages of 
$\sim 0.5-1.5\,\textrm{Gyr}$ is a factor of $\sim 2$, smaller than 
the change in the $K$-band ratio.
Calculating the mass from $M_{H}$ also has the advantage that
the contribution of the power-law emission is naturally lower in the 
$H$-band than in $K$-band, so the uncertainty introduced 
due to our subtraction of the power-law component is smaller.
A final reason for the use of $H$-band in deriving the stellar masses 
is that for the highest-redshift ($z>3$) SMGs in our sample, 
the longest-wavelength IRAC observations (8.0\,$\micron$) 
do not quite sample rest-frame $K$, so calculations of the 
absolute $K$ magnitude ($M_{K}$) for these sources are not
constrained by the available data without extrapolation. 

In \S\ref{sec:sfhistory}, we explained our rationale for adopting the stellar
mass calculation method of \citet{borys05}, which uses a single 
$L/M$ ratio for the entire sample of SMGs for a given star formation history.
We obtain this $L_{H}/M_{\star}$ ratio for each star formation history (IB and CSF)
by finding the average of the best-fit model ages over the SMG sample 
for the fits of that star formation
history and taking the $L/M$ ratio for that star formation history
corresponding to that average age. Thus, for the BC03 models we use
$L_{H}/M_{\star} = 5.8\,\lsun\,\msun^{-1}$ for the CSF
 history and $L_{H}/M_{\star} = 7.9\,\lsun\,\msun^{-1}$ 
for the IB history, corresponding to average sample ages of 
1.5\,Gyr and 140\,Myr, respectively.  From the M05 
stellar models we derive an average age and $L/M$ ratio of
1.6\,Gyr and $L_{H}/M_{\star} = 7.9\,\lsun\,\msun^{-1}$ for the CSF
history, and 120\,Myr and $L_{H}/M_{\star} = 8.5\,\lsun\,\msun^{-1}$ for the IB
history.  

We calculate the stellar mass for each SMG for each of
the CSF and IB star formation histories 
using the $L_{H}/M_{\star}$ ratio for that history and $M_{H}$
from the normalization of the SED fit of that 
star formation history, then average the masses obtained 
from the IB and CSF histories 
to obtain our final stellar mass estimate for each SMG.
We estimate the uncertainty in the stellar masses by 
dividing in half the difference in the masses resulting from
the CSF and IB star formation histories, since these values tend 
to represent the maximum and minimum values of the $L_H/M_{\star}$ 
ratio among the various simple star formation histories we examined 
(see Figure~\ref{fig:lmratio_vs_time}).  
We thus attempt to represent the systematic effects
of star formation history and $L_H/M_{\star}$ ratio in our 
uncertainty estimates, which are much more significant than the random 
errors in the model fitting due to the photometric uncertainties
\citep[see, e.g.,][]{muzzin09}.   However, we caution
that myriad systematic errors are still possible for our mass estimates 
based on the assumptions we have made,
as we discuss in the next section.

\subsection{Sources of Systematic Uncertainty in Median $M_{\star}$}\label{sec:sys_unc}

Although the median stellar mass of our SMG sample is the best-constrained 
quantity we derive in our study, the median stellar mass still has associated
random and systematic uncertainties which can be significant and must be
mentioned.  We have estimated the random error in our median stellar mass
by randomly perturbing the observed photometry of our sample within its
associated error bars and carrying out our SED fitting and mass estimation
procedure on the perturbed data for the galaxies.  By analyzing the statistics
over many trials (where each trial produces a median stellar mass for the sample),
we find that the standard deviation of the median stellar mass, which we call
the random error in the median of our sample, is 11\%.  
Because the random error associated with our SED fitting is
only $\sim 10$\%, the overall uncertainty in our median
stellar mass is dominated by systematic uncertainty.  On the whole, we expect
the dominant systematic uncertainty arises from our use of 
stellar population synthesis models to obtain a $L/M$ ratio; as this 
uncertainty is difficult to quantify and affects nearly all studies 
of the stellar mass in high-redshift galaxies similarly, we refrain from discussing it at length. 
Rather, we restrict ourselves to a brief summary of the more concrete systematic
effects and their impact on our median stellar mass, noting that
many of the sources of uncertainty have been explored in detail
elsewhere \citep[e.g.,][]{papovich01,vanderwel06,conroy09,muzzin09}. 
We further note that, in principle, systematic errors in the stellar mass calculations
could occur within our sample between the different sky fields because
some of the fields have fewer optical photometric data points to 
constrain the SED models from which derive the ages and $M/L$ ratios.  However, in practice, since our
aim is to constrain the average stellar mass over the SMG sample and not 
individual galaxy masses, this particular source of systematic error is not
significant: tests varying the number of optical photometric data points used in fitting
the GOODS-N subsample of SMGs (which has the most photometric data of
all the fields in our sample) reveal variation of no more than 0.1\,dex in the
mean and median stellar mass.  

Relatively minor contributions ($\sim 10-20$\% individually) to the systematic uncertainty
in our median stellar mass arise from our metallicity and extinction law
assumptions and our removal of the near-IR continuum excess in fitting
the SEDs of our sample of SMGs.  As shown by \citet{muzzin09} and \citet{conroy09},
the assumption of a single metallicity in our fitting introduces uncertainties
in the stellar masses of 10--20\%.  We have estimated the uncertainty in the 
median stellar mass resulting from use of the 
\citet{calzetti00} extinction law for starbursts to be also 10--20\% by 
fitting stellar population models to the SEDs of our sample
assuming the extinction laws for the Milky Way \citep{fitzpatrick86} and the 
Small Magellanic Cloud \citep{prevot84,bouchet85}.  
We note that our simplistic treatment of the extinction within SMGs as a 
uniform dust screen also affects our stellar mass estimates in a
way which is difficult to quantify;
yet, as there are essentially no constraints on the distribution 
of dust and star formation within SMGs, it is difficult to justify the use
of more complex extinction models, such as age-dependent
extinction \citep[e.g.,][]{poggianti00}.  
The systematic uncertainty in our average stellar mass estimate associated with
our near-IR continuum removal procedure arises from the index of the 
PL subtracted from the observed SED to obtain the stellar $H$-band luminosity.
We have estimated this uncertainty as 10--15\% by comparing the stellar masses of
the individual galaxies in our SMG sample resulting
from the assumption of PL indices $\alpha = 2$ and $\alpha=3$.   

The systematic uncertainty in our typical stellar mass estimate from
our choice of IMF is considerably larger than those associated with metallicity, 
extinction, and choice of PL index.  The choice of a particular function
effectively introduces a scaling factor into the mass estimates: for example,
if we instead chose to use the Salpeter IMF with the BC03 and M05 stellar
population models rather than the \citet{chabrier03} and \citet{kroupa01} IMFs,
the median stellar mass would be higher by a factor of 1.6 (Kroupa) and 1.8 (Chabrier).  
If we assume the systematic uncertainties arising from the choice of IMF,
metallicity, extinction law, and PL subtraction are all independent (which may not be correct)
and add them in quadrature, we obtain a conservative estimate of the 
systematic uncertainty in our typical stellar mass of a factor of 1.6--1.8, which
is clearly dominated by the IMF systematic.  Including the uncertainty due
to the stellar population models 
will obviously increase the overall systematic error estimate.

We caution that a final source of systematic uncertainty which
is not easy to quantify with currently existing observational data 
may significantly affect our stellar mass estimates.  Should a 
large fraction of the stellar content of SMGs lie within or behind
the heavily obscured star formation regions, we are unable to
trace its mass since we do not receive its light.  In principle,
dynamical mass estimates can help assess the amount of mass
missing from our estimates, but the currently available dynamical
masses for SMGs come from CO or H$\alpha$ observations which
may not trace the bulk of the stellar mass.  In the future, when
precise dynamical masses for our sample are obtained from rest-frame
near-IR (i.e., stellar) light, we will be able to fully address this issue.
Until then, our stellar masses cannot include this unknowable, but potentially significant,
fraction of the stellar mass in SMGs.  How much mass
is missing from our estimate of the median will depend on how the dust is
distributed throughout SMGs, which we discuss further in the next section.

\section{RESULTS}\label{sec:mstar_sec}

In Table~\ref{tab:massresults}, we provide for  
our sample of 65 spectroscopically-identified SMGs the de-reddened 
values of $M_{H}$ and the $M_{\star}$ calculated as described in 
\S\ref{sec:mass_est_proc}, as well as averaged $V$-band
extinctions and the 8.0\,$\micron$ PL fraction resulting from our  
stellar population synthesis model plus PL fitting for the M05 stellar
models.  The results for the BC03 models are largely 
similar, though with a scaling factor in the stellar masses,
so we have chosen not to tabulate those results here.  As a 
brief summary of the differences between the BC03 and M05 results,
we find that the absolute $H$-band magnitudes resulting from the 
use of the BC03 models are 0.03\,mag fainter, on average, than those
resulting from the M05 models; the $V$-band extinctions are equal, on average, 
between the BC03 and M05 models for the IB star
formation history but the extinctions for the CSF star formation
history are 0.4\,mag higher systematically in the BC03 results; finally,
the stellar masses resulting from the BC03 models are systematically
25\% higher than those obtained with the M05 models, although we
note that a Chabrier IMF has been used for the BC03 results
while a Kroupa IMF has been used for the M05 results.
We discuss the results contained in Table~\ref{tab:massresults}, some
of their implications, and some caveats in the sections below, providing
the quantities derived from the BC03 models only when significantly different. For our
discussion we separate the $z<1.5$ SMGs from the higher-$z$ galaxies
because our focus in this paper is on the high-redshift SMGs which
dominate our sample, cover a smaller epoch of cosmic time, and are more 
important in a cosmological sense in that they contribute more significantly to
the cosmic star formation rate density ($z\sim 2$ SMGs with $S_{850} > 4\,\textrm{mJy}$
contribute $\sim25$\% of the overall star formation rate density at that redshift, while
similarly bright $z \lesssim 1$ SMGs contribute $\sim10$\%; C05; Wardlow et al.\ 2011).  
Moreover, the low-$z$ SMGs are known to differ 
somewhat from the high-$z$ SMGs in that they are
typically less far-IR-luminous (C05), and thus may have 
different characteristics from the high-$z$ sub-sample overall.

\subsection{Contribution of Non-Stellar Near-IR Emission in SMGs}\label{sec:pl_results}

\begin{figure}
\begin{center}
\includegraphics[width=1.0\columnwidth,trim=40pt 10pt 10pt 10pt,clip]{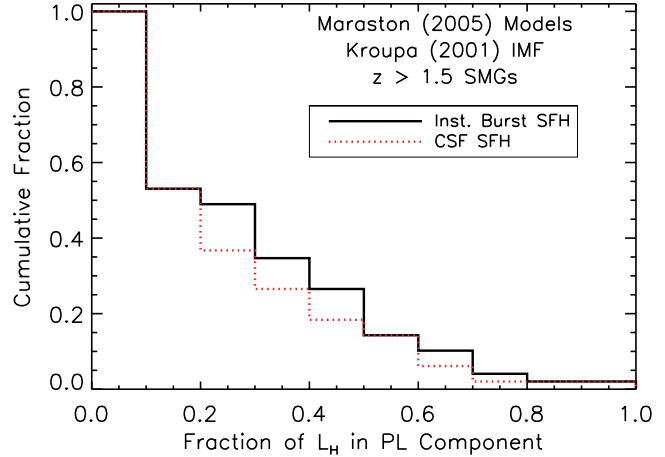}
\caption{Cumulative distribution of the fraction of rest-frame $H$-band luminosity
in the power-law component for $z>1.5$ SMGs, based on the best-fit power law fraction
at observed-frame 8.0\,$\micron$, using 
the M05 stellar population models.  Approximately half the SMG sample has less than
10\% contribution to the rest-frame $H$ luminosity from the power-law component,
and 11\% of the sample has 
more than 50\% of the $H$-band emission originating 
from the power-law component component.}\label{fig:powlaw_dist}
\end{center}
\end{figure}

Through fitting the combination of stellar population models and a PL 
to the observed SED of our sample of SMGs, we have placed constraints on
the fraction of the observed-frame 8.0\,$\micron$ emission of each
SMG which is contributed by a PL component, which we tabulate in Table~\ref{tab:massresults}.  
To make our results easier to interpret, we have converted this
best-fit PL fraction at observed-frame 8.0\,$\micron$ into
the fraction of the total rest-frame $H$-band luminosity 
($L_{H}$; approximately the location of the peak of the stellar emission)
of each SMG in our sample contributed by its PL component.
In Figure~\ref{fig:powlaw_dist} we show the cumulative distribution of 
the calculated fraction of rest-frame $H$ luminosity 
in the PL component for the $z>1.5$ portion of our sample
for both the IB and CSF star formation histories 
using the M05 stellar population models.  The
distributions of inferred PL fraction are quite similar
for the different star formation histories; for each, approximately
half of our $z>1.5$ sample have essentially negligible contributions
to their total $L_{H}$ from a PL component.  Of our entire spectroscopic
SMG sample, 69\% (79\%) of the galaxies have non-zero fractions
of $L_{H}$ for the CSF (IB) star formation history; yet for only 11\% 
(for both star formation histories) is 
the fraction of $L_{H}$ in the PL component greater than 0.5.  For reference,
and warning, we list in Table~\ref{tab:smg_agns} those SMGs for which the fraction
of $L_{H}$ in the PL component is greater than 0.5. 
Clearly, the majority of our SMG sample are stellar-dominated in
rest-frame $H$, although for much of the sample the stellar mass 
will be somewhat overestimated (by a median value of 10\%, and
a mean value of 25\%) if we assume that all of the near-IR 
light is stellar.  While the amount by which the stellar masses are 
overestimated lies well within the errors of our present study (see
\S\ref{sec:sys_unc}), future 
studies which aim to determine $M_{\star}$ more precisely for individual
SMGs will need to remove the non-stellar contribution to the near-IR light.
 
Interestingly, one of the near-IR PL-dominated SMGs listed in 
Table~\ref{tab:smg_agns} also displays a rest-frame UV excess.  At least three
additional galaxies in our sample also appear to have a UV excess, which
are noted in Table~\ref{tab:massresults}; we note that because the detection of
a UV excess requires a reported observation in the observed-frame $U$ band, 
which most of sample lack, we are prevented from accurately determining the frequency
with which UV excesses occur in our sample.  Each of the four
UV-excess SMGs have significant PL contributions in the near-IR,
even though three of the four galaxies do not qualify as PL-dominated. For two 
of these SMGs (SMM\,J123632.61+620800.1 and SMM\,J123635.59+621424.1) 
AGN spectral identifications in the observed-frame optical (C05)
and X-ray \citep{alexander05} suggest that a type-1 AGN could be
the origin of the UV excess; high-resolution imaging of these galaxies in
the observed-frame $U$-band showing a compact morphology would
lend additional support to such a hypothesis.  However, the other two 
UV-excess SMGs (SMM\,J123553.26+621337.7 and SMM\,J123555.14+620901.7)
have no AGN signatures in their observed-frame optical
spectra, so attributing the UV excess to a type-1 AGN is problematic.
For these SMGs, a possible explanation for the UV excess is that a
simple uniform dust screen is simply an inappropriate model of the
stellar extinction across the entire galaxy, as \citet{poggianti01}
show for luminous IR galaxies displaying e(a) spectra.  
For example, in these particular SMGs some portion of the total star 
formation may be relatively unobscured and contribute some UV flux 
to the SED, while the dominant star formation activity is heavily
obscured.  Another possibility is that the UV excess is a result of
scattering of obscured AGN light, similar to the origin of the far-UV flux
in the local Seyfert 2 galaxy NGC\,1068 \citep[e.g.,][]{code93,grimes99}.
High-resolution $U$-band imaging of these SMGs will also be 
important to determine the origin of the UV excess.

\subsection{Near-IR Stellar Luminosity of SMGs}\label{sec:stellar_lum_section}

As a result of our separation of the PL component from the
stellar component in the observed SEDs of our sample of SMGs, 
we have obtained more representative estimates of the 
rest-frame near-IR stellar luminosity for SMGs.  The absolute $H$
magnitudes resulting from our SED fitting procedure are much less
dependent on star formation history than the stellar masses, as the
absolute magnitudes are calculated
from the photometric data point nearest the rest-frame $H$ band
by applying a (small) $K$-correction interpolated from the best-fit
stellar population model.  Thus, the absolute magnitudes of individual SMGs
listed in Table~\ref{tab:massresults} are much better constrained than 
the stellar masses listed for individual galaxies.
Since we fitted both IB and CSF star formation histories to the 
observed SMG SEDs,  we obtain multiple estimates of
$M_{H}$; to obtain a single $M_{H}$ for each SMG for our present discussion 
and for Table~\ref{tab:massresults}, we average the $M_{H}$ values 
resulting from the fits of the IB and CSF star formation histories. 
We estimate the uncertainty as half of the difference between the 
IB and CSF values; by doing so we take into account the 
uncertainty in the $K$-correction determined from the SED models of
different star formation history, although we anticipate it is small.  

The median of the stellar absolute 
$H$ magnitude determined in this way for the $z>1.5$ SMGs in 
our sample, including the upper limits for galaxies which 
had fewer than 3 detections in their SEDs, 
is $\langle M_{H} \rangle = -25.7 \pm 0.2$ with standard deviation
$\sigma_{M_{H}} = 1.1$ for the M05 models; we note that the median $M_{H}$
found using the BC03 models is only 0.02\,mag fainter.  
When de-reddened by the median value of $A_{V}$, the median absolute
$H$-band magnitude for becomes $\langle M_{H} \rangle = -26.0$.
For the $z<1.5$ SMGs, we find a median, uncorrected for reddening, of
$\langle M_{H} \rangle = -25.3 \pm 0.7$ with standard deviation
$\sigma_{M_{H}} = 1.7$ for the M05 models, and when de-reddened
by the median $A_{V}$, the median becomes $\langle M_{H} \rangle = -25.6$.
There is thus no statistically significant difference in the typical
$M_{H}$ between the low- and high-$z$ sub-samples.  
Both the low- and high-$z$ sub-samples of SMGs have typical
stellar luminosities which are greater than local $L^{\ast}$ galaxies
\citep[$M^{\ast}_{H} = -23.7$;][]{cole01},
which suggests that the stellar populations of SMGs are younger, 
that typical SMGs are more massive than local $L^{\ast}$ galaxies,
or some combination of the two.  Given that we observe  
these SMGs at high redshift, it is highly likely that their stellar 
populations are younger than those of old local $L^{\ast}$ galaxies,
but we should compare the stellar masses of SMGs to those of $L^{\ast}$
galaxies directly to examine the effect of mass.

\subsection{Stellar Mass Results for SMGs}\label{sec:mstar_results}

\begin{figure}
\begin{center}
\includegraphics[width=1.0\columnwidth,trim=2pt 2pt 7pt 0pt,clip]{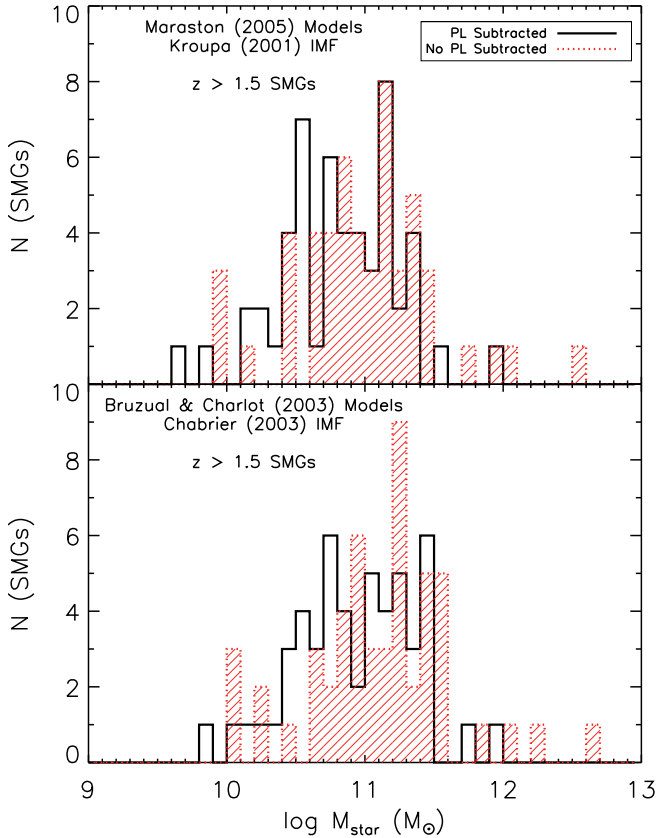}
\caption{Histograms of stellar mass
($M_{\star}$) for C05 sample of SMGs using the stellar population
synthesis models of M05 (top) and BC03 (bottom).  The filled histogram
indicates the $M_{\star}$ distribution when no power-law component is
subtracted as part of the SED fitting; in those distributions a slight shift to higher 
$M_{\star}$ can be observed.  The distributions span approximately an order
of magnitude, and the power-law-subtracted distributions 
have medians $(7.2 \pm 1.2) \times 10^{10}\,\msun$ (M05 models)
and $(9.0 \pm 1.7) \times 10^{10}\,\msun$ (BC03 models). }\label{fig:masshist}
\end{center}
\end{figure}

As mentioned in \S\ref{sec:mass_est_proc}, we have chosen a stellar mass
estimation procedure which constrains the average over the entire 
sample at the expense of accuracy in the mass of the individual galaxies.
As a result, although we list stellar masses for individual SMGs in
Table~\ref{tab:massresults} for readers who wish to use them in spite of
their uncertainty (factor of $\sim 2$; see \S\ref{sec:mass_est_proc}), 
we do not wish to focus on results in an individual sense.  We 
find it more appropriate to discuss the average (median/mean) mass
of the sample and its range, which are not as sensitive to the star formation
history uncertainty due to our large sample size. 
In Figure~\ref{fig:masshist} we show the distribution of 
$M_{\star}$ for the $z>1.5$ SMGs in 
our sample for both the M05 and the BC03 models to visually 
display the range in stellar mass of our sample.  We caution that
we have included in the histograms the stellar masses even for 
SMGs whose rest-frame optical and/or near-IR flux appeared to be dominated
by a PL component prior to its removal, and remind the reader that
the M05 stellar mass estimates use a different IMF \citep[that of][]{kroupa01} 
from the BC03 stellar mass estimates \citep[that of][]{chabrier03}. 
While the distributions for the M05 and BC03 models have 
similar widths, $\sigma \sim 0.4\,\textrm{dex}$, 
the distribution of BC03 stellar masses is shifted relative to 
the distribution of M05 masses such that, even when corrected to the
same IMF, the M05 masses are systematically $\sim 20$\% lower 
than the BC03 masses on average.  We attribute at least some of the
difference in the masses estimated using the M05 and BC03 models to
the effects on the $L_{H}/M_{\star}$ 
ratio of TP-AGB stars included in the models
of M05.  At the average age of SMGs in the CSF
model for M05, 1.6\,Gyr, the $L_{H}/M_{\star}$ ratio is
$\sim 40$\% higher than that for the BC03 models at 
the average SMG age, 1.5\,Gyr.
We adopt the masses from the M05 models for the 
remainder of our discussion since the inclusion of the TP-AGB phase
in stellar evolution in those models makes them more likely to be 
representative of the actual $L/M$ ratio.

The median of the stellar mass distribution for the $z>1.5$ SMG sample is  
$\langle M_{\star} \rangle = (7.2\pm 1.2) \times 10^{10}\,\msun$ 
for the M05 models [$\langle M_{\star} \rangle = (9.0 \pm 1.7) \times 10^{10}\,\msun$
for the BC03 models].  The inter-quartile range (25th -- 75th percentile) is 
$3.3\times 10^{10} - 1.3\times 10^{11}\,\msun$, 
suggesting that the $z>1.5$ SMGs span a range of approximately an order 
of magnitude in stellar mass.  The $z<1.5$ SMGs in our sample
seem slightly less massive on average than the
high-$z$ SMGs, having a median stellar mass of 
$\langle M_{\star} \rangle = (4.4 \pm 1.7) \times 10^{10}\,\msun$ for the
M05 models; however, the difference in median mass
between the low-$z$ and high-$z$ SMGs is only $\sim 2\sigma$, and
so not statistically significant.  We note that the new median $M_{\star}$ we obtain for 
the $z > 1.5$ SMGs lies between that 
of model $z \sim 2$ SMGs in the semi-analytic model examined in 
\citet[][$M_{\star} = 2.1 \times 10^{10}\,\msun$]{swinbank08} and
the model SMGs in the hydrodynamical simulations of 
\citet[][$M_{\star} = 2.7 \times 10^{11}\,\msun$]{dave10}; thus
our new estimate does not favor one set of theoretical models
over the other.  We also note that the median stellar mass we obtain
here for $z > 1.5$ SMGs does not qualitatively change the 
conclusion of \citet{alexander08} that the majority of
SMGs fall a factor of 3--5 below comparably massive
local galaxies on the $M_{BH}-M_{\star}$ relation; thus, 
any lag in the growth of the central black holes
relative to the host galaxy is modest in SMGs.

The typical stellar mass of SMGs found in previous observational studies span
an order of magnitude; our new result is intermediate between
the estimate of \citet[][$M_{\star} \sim 3 \times 10^{10}\,\msun$]{smail04}, 
who lacked rest-frame near-IR data for their
analysis and thus could only weakly constrain the stellar mass, and
the results of previous studies which used rest-frame 
near-IR data from \emph{Spitzer}-IRAC \citep{borys05,dye08,michalowski10}. Although 
lower, our median $M_{\star}$ is marginally consistent with that obtained by  
\citet[][$\langle M_{\star} \rangle = 3.3 \pm 1.2 \times 10^{11}\,\msun$]{borys05}, 
whose SMG sample is contained fully within ours, falling just outside
their $\pm 2\sigma$ uncertainty interval.   A better comparison
to \citeauthor{borys05} is to compute the median
$M_{\star}$ for only the subsample of SMGs which 
were also studied by those authors; when we do so, we obtain
$\langle M_{\star} \rangle = (1.3\pm 0.3) \times 10^{11}\,\msun$, which
falls within the $\pm 2\sigma$ uncertainty interval of \citeauthor{borys05}. 
We note that since the stellar population models we have used for our
study are not available for the IMF used by \citeauthor{borys05}, that
of \citet{millerscaloimf}, we cannot make a precise comparison between
the two results; however, the smaller contribution of low mass stars
in the \citet{millerscaloimf} IMF relative to the Kroupa IMF suggests that the 
discrepancy in the median between the two studies may be somewhat larger 
under the \citet{millerscaloimf} IMF.   To compare our new result to those of
\citet{dye08} and \citet{michalowski10}, who use a Salpeter IMF, we 
have also carried out our SED fits with a Salpeter IMF and find that our
median mass for $z > 1.5$ SMGs increases to $1.3\times 10^{11}\,\msun$. 
Despite the increase resulting from the IMF conversion, our typical stellar mass
is still significantly smaller than that found by
\citet[][$\langle M_{\star} \rangle = 6.3_{-1.3}^{+1.6}\times 10^{11}\,\msun$]{dye08}, 
falling outside their $\pm 3\sigma$ uncertainty interval.  \citet{michalowski10} 
obtain a median stellar mass of $\langle M_{\star} \rangle = 3.5\times 10^{11}\,\msun$,
a factor of 2.7 larger than our median $M_{\star}$; the significance of
the difference is unclear since \citeauthor{michalowski10} provide no uncertainty
estimates for their individual or median stellar masses.

While the fact that we have used different stellar population models 
(e.g., \citealt{borys05}\ use the stellar population models of \citealt{bc93}) 
likely plays a role in our systematically lower stellar mass
estimates, the uncertainties associated with
the different models are insufficient 
to fully account for the discrepancies.   One important factor
underlying our smaller stellar mass estimates is that we
do not assume that the near-IR light from SMGs is entirely stellar in origin, 
unlike the studies of \citet{borys05} and \citet{dye08}.
Rather, we have subtracted the non-stellar near-IR excess from the 
observed photometry and used $M_{H}$ to calculate the stellar mass instead of
$M_{K}$ since the contribution of the power-law emission is lower in the 
$H$-band than in $K$-band.   
If we instead estimate $M_{\star}$ for our sample using rest-frame 
$K$-band absolute magnitudes and \emph{do not} subtract the near-IR excess,
the median mass for our sample is $\langle M_{\star} \rangle = 1.3 \times 10^{11}\,\msun$
under the M05 models and Kroupa IMF,
a factor of $\sim 2$ higher, reflecting the impact of the near-IR continuum excess
on the stellar masses in an average sense.   
We thus expect that the previous studies in which the near-IR continuum excess 
was not subtracted prior to stellar mass calculation have
systematically overestimated the stellar masses of SMGs with near-IR
excesses.  

In \S\ref{sec:stellar_lum_section} above we noted that typical 
SMGs have brighter stellar $M_{H}$ than local $L^{\ast}$ galaxies, 
which could be due to a higher stellar mass in SMGs or younger stellar
population ages.  It is thus notable that we find that
the typical stellar mass of high-$z$ SMGs is a factor 
of 2 \emph{below} the typical stellar mass of local $L^{\ast}$ galaxies 
\citep[$m^{\ast} \sim 1.4\times 10^{11}\,\msun$;][]{cole01}.  Thus, the
difference in stellar $M_{H}$ between SMGs and local $L^{\ast}$ galaxies
can be attributed to the younger stellar population ages of SMGs.
We note, though, that given the systematic errors in our mass estimates, 
which we discussed earlier in \S\ref{sec:sys_unc}, 
we consider the typical stellar mass of high-$z$ SMGs
to be roughly consistent with $m^{\ast}$.  
We will return to the comparison of 
SMGs with local galaxies in \S\ref{sec:gasfrac}.

\subsection{Average Extinction of the Stellar Populations of SMGs}

\begin{figure}
\begin{center}
\includegraphics[width=\columnwidth,trim=18pt 15pt 10pt 10pt,clip]{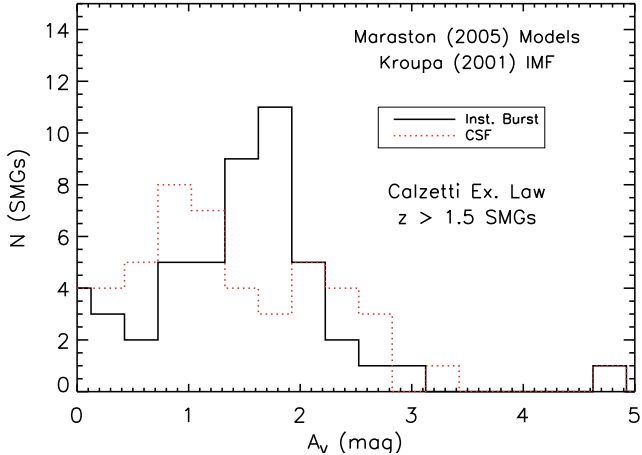}
\caption{Histograms of
$A_{V}$ for C05 sample of SMGs using the stellar population
synthesis models of M05 for two different choices of star formation
history, assuming the \citet{calzetti00} extinction
law for starburst galaxies. The total near-IR light we observe from
SMGs is only modestly extincted, having median $\langle A_{V} \rangle = 1.2-1.5$,
depending on the choice of star formation history.}\label{fig:redd_dist}
\end{center}
\end{figure}

Our stellar population model fitting, which assumes a simple
dust screen model, also provides estimates
of the typical $V$-band extinction for our sample of SMGs.  
Because the SMGs are unresolved in their observed-frame optical--IRAC photometry, we 
regard these extinctions as averages across the entire galaxy; the true
obscuration probably varies with location.  However, since the
extinction estimates are correlated with stellar population
age and star formation history, which we could not firmly constrain
for individual SMGs, we once again consider the average
and range over the entire sample to be the best-constrained quantities.
We show the distribution of the fitted extinctions for both IB and CSF
star formation histories for $z> 1.5$ SMGs in Figure~\ref{fig:redd_dist},
which illustrate both the likely range of $A_{V}$ over the sample and the
effect of star formation history on the extinction estimates.  
The distributions for both star formation histories indicate
that the galaxy-averaged optical and near-infrared 
light that we detect from high-$z$ SMGs is typically moderately obscured
($0.5 < A_{V} < 1.5$), with medians of 
$\langle A_{V,burst} \rangle = 1.5\pm 0.1$\,mag and 
$\langle A_{V,CSF} \rangle = 1.2\pm 0.2$\,mag and standard deviations
of 0.8 and 1.0\,mag, respectively.  The extinction
in the $z<1.5$ sub-sample spans a similar range, with medians 
$\langle A_{V,burst} \rangle = 1.6\pm 0.2$\,mag and
$\langle A_{V,CSF} \rangle = 1.8\pm 0.3$\,mag and standard deviations
of 0.8 and 0.7\,mag.  To give the reader a sense of the difference in 
$A_{V}$ between star formation histories for individual SMGs, the values 
we list in Table~\ref{tab:massresults} are the average of the $A_{V}$'s
found for the CSF and IB star formation histories, and the errors are 
taken to be half the difference between the two. 
The typically modest visual extinctions we obtain are consistent with
those estimated from the Balmer decrement \citep{takata06}, 
but contrast sharply with the extreme IR luminosities of SMGs, 
which imply powerful energy sources
hidden behind copious dust, and with the typical visual extinction
derived from the optical depth of the 9.7\,$\micron$ silicate absorption
feature in the mid-IR spectra of SMGs \citep[$A_{V} \sim 6$;][]{karin09}.
However, we may reconcile the optical/near-IR $A_{V}$ estimate with the mid-IR
estimate by noting that the mid-IR light is less sensitive to dust extinction
so we receive emission from more obscured regions at mid-IR wavelengths 
than at optical wavelengths, thus increasing the average $A_{V}$ we
derive.  

The discrepancy between optical and mid-IR estimates of $A_{V}$
highlights the difficulty in quantifying the stellar mass behind heavily
obscured star formation complexes by means of an UV--near-IR SED fitting 
analysis, and again we caution that such
mass may be missing from our estimates in this paper.  How much the 
mass has been underestimated depends critically on the dust distribution
in SMGs, which remains poorly known.  If the star forming clouds and 
heavy extinction are patchy and distributed over the galaxy, then older stars
will eventually diffuse out of the heavily obscured regions (perhaps on 
Milky-Way-type time-scales of $\sim 100\,\textrm{Myr}$).  In this case, we would
be able to observe the bulk of the stellar mass, except in the earliest
formation stages of the galaxy.  On the other hand, if the entire galaxy
is enshrouded in dust, and the older stars never diffuse out to 
regions of lower extinction, a large fraction of the stellar mass could
be hidden.  While comparisons with dynamical mass estimates of 
SMGs (see \S\ref{sec:cocompsec}) suggest that a large fraction of 
the stellar mass is not missing from our estimates, the dynamical 
masses are derived using a tracer (gas) which may not have the same
spatial distribution as the stars and may not be representative of 
entire galaxies.

\section{DISCUSSION}\label{sec:mstar_discussion}

In this section we combine the stellar luminosities and typical stellar mass 
we have derived for our SMG sample with various mass, AGN, and
star formation indicators to extract their implications for the  
nature of SMGs and their future evolutionary path.  
We first perform a small, but independent, 
check of our typical stellar mass estimate to evaluate its plausibility before
moving on to perform comparisons of SMGs to other populations of 
high-$z$ galaxies.  We will also discuss the
origin of the non-stellar near-IR continuum emission.  Finally, we 
will use constraints from the fading over time of stellar population models 
to predict the $z=0$ descendants of SMGs.

\subsection{Comparison to Dynamical Mass from CO Observations: 
Testing $M_{\star}$}\label{sec:cocompsec}

The various different systematic uncertainties associated with 
stellar mass determinations, and the different results which have
been obtained in the literature strongly highlight the need for
independent checks on stellar mass estimates.
In principle, the best check on any stellar mass estimate is to compare 
$M_{\star}$ to the dynamical mass, $M_{dyn}$.  However,
\citet{grevesmg} warn that the dynamical masses of SMGs must be 
treated with caution, because $M_{dyn}$ for SMGs (and other galaxies)
is usually calculated assuming the system is in dynamical 
equilibrium and that we know the velocity field and mass distribution,
whereas SMGs may be the product of a merging gas-rich system 
\citep[e.g.,][]{smail98,smail04,tecza04,swinbank06,tacconi08}
which is not in dynamical equilibrium.  
If the system is not in dynamical equilibrium, the mass distribution assumed
is incorrect, or if the observed CO line is not a good tracer of the system 
dynamics or mass, $M_{dyn}$ will be different.  Moreover, high-resolution
millimeter observations by \citet{tacconi06,tacconi08} and \citet{bothwell10} 
suggest that the CO-emitting region 
($R \sim 1-2\,\textrm{kpc}$) in SMGs is frequently smaller than the overall
stellar mass and star formation distribution \citep{chapman04,biggs06,swinbank10}, 
so the dynamical masses from CO observations are often lower limits.
A further complication is that the dynamical mass is not 
exclusively a baryonic mass: since a galaxy's dark matter halo contributes
to the gravitational potential and thus to the velocity field, the galaxy's
mass derived from the velocity field will contain some contribution
from dark matter.  The fractional contribution of dark matter to the 
available CO dynamical mass estimates is currently unknown; however,
the likely descendants of SMGs in the local universe
(i.e., galaxies of equal or greater stellar mass) have been
observed to be baryon-dominated on scales similar to the CO-emitting
region in SMGs \citep[e.g.,][]{gerhard01,cappellari06,kassin06,williams09}.

In spite of the uncertainties, we use the 16 SMGs with CO observations in our sample 
to carry out the best test available of the stellar masses we have derived.  
We list the molecular gas results and dynamical mass estimates for the 16 SMGs 
which have published CO observations in Table~\ref{tab:co_sources}, noting 
that only the CO-detected SMGs have estimates of $M_{dyn}$.
\citet{tacconi08} have carried out a comparison of gas, stellar, and 
dynamical mass for four of the SMGs in our sample already, in order to constrain the IMF
and CO-to-$\hh$ conversion factor for high-$z$ galaxies, and found that
the stellar masses of two of the SMGs exceeded the dynamical mass.  
However, we are more concerned here in determining if the typical 
stellar mass of SMGs is consistent with the typical dynamical mass, 
since our stellar masses for individual SMGs are highly uncertain. 
The median dynamical mass for SMGs in our sample with 
CO observations is $\langle M_{dyn} \rangle = 2.3 \times 10^{11}\,\msun$ and
thus consistent with the sum of the median stellar mass ($M_{\star} = 7.2\times 10^{10}\,\msun$)
and the median molecular gas mass [$M(\hh) = 3.0\times 10^{10}\,\msun$]
of the SMGs with CO observations \citep[which is identical to the median of the 
combined samples of][]{grevesmg,tacconi06,tacconi08}.
Thus, assuming the median $M_{dyn}$ and $M(\hh)$ are representative
of the whole sample, we conclude that our median stellar mass estimate
is generally consistent with the published dynamical masses of SMGs.

\subsection{The Origin of the Near-IR Continuum Excess in SMGs}

Although the majority of our SMG sample does not contain a dominant near-IR PL
component, we nevertheless wish to understand the origin of the near-IR excess observed in
a significant fraction of the galaxies in our sample.  To test the
origin we need to distinguish the near-IR emission from the galaxies' nuclei
from that which follows the obscured star formation \citep[which may be more
distributed; see, e.g.,][]{chapman04,biggs06,karin09}.  Yet, because
high-$z$ SMGs are unresolved in our IRAC images, we cannot directly connect
the PL emission to the galaxies' nuclei or star formation regions; 
we must find some other means of 
discriminating between the central SMBH and the host galaxy as 
the source of the near-IR excess.  One way to do so is to examine other
AGN indicators in the SMGs: if the origin of the excess is AGN, we would 
expect that the presence and/or luminosity of a significant PL component 
correlates with other AGN signatures, unless the AGN is highly obscured 
and Compton-thick \citep[which does not appear to be the case for the 
majority of SMGs; see][]{alexander05}.  

\begin{figure}
\begin{center}
\includegraphics[width=\columnwidth,trim=32pt 15pt 10pt 20pt,clip]{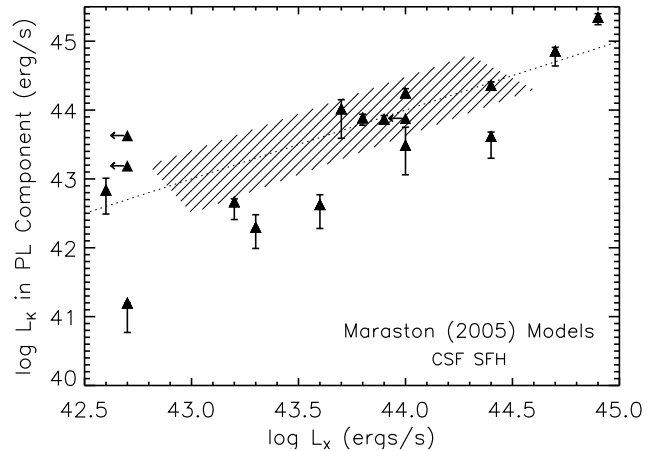}
\caption{Rest-frame $K$-band luminosity in the PL component as a function of
extinction-corrected X-ray luminosity for the SMGs which have X-ray observations
in the literature. The dotted line represents $L_{K}(\textrm{PL Component}) = L_{X}$.
The shaded area shows the region in which local AGN fall.
With the exception of 2 SMGs which may be Compton-thick AGN,
and 3 SMGs which have no near-IR PL component, $L_{X}$ is clearly correlated
with the $K$-band luminosity of the non-stellar continuum excess; the correlation
suggests that AGN are the origin of the near-IR continuum 
excess in SMGs.}\label{fig:lk_vs_lx}
\end{center}
\end{figure}

The hard X-ray luminosity of a galaxy is regarded as one of the 
best tracers of SMBH activity for Compton-thin nuclei, since the hard
X-rays are less susceptible to the heavy obscuration thought to enshroud
a central SMBH.  Moreover, numerous authors have shown that the hard X-ray
luminosity of QSOs and Seyfert 1 galaxies is correlated with the nuclear
near-IR luminosity \citep[e.g.,][]{carleton87,alonsoherrero97,mushotzky08},
who frequently conclude that the non-stellar near-IR (e.g., $K$, $L$, and $M$ bands)
emission originates from hot dust heated by the AGN. 
23 SMGs in our sample have rest-frame hard X-ray 
data available in the literature \citep{alexander05,alexander08,lefloch07}.
For these galaxies we can most directly test if the near-IR excess
is associated with an AGN, by checking if the hard X-ray luminosity 
correlates with the luminosity in the near-IR PL component.  
Although the most significant correlations in local AGN samples 
between near-IR and X-ray emission occur in $L$ and $M$ bands due to stellar
contamination in bluer bands, our IRAC data cannot constrain bands redward
of rest-frame $K$ for most of our sample.  Thus, to test for
a correlation, we have calculated the rest-frame $K$-band luminosity
contained within the PL component for each of the 23 SMGs with X-ray data
(in addition to computing the same for rest-frame $H$-band in 
\S\ref{sec:pl_results}). 
In Figure~\ref{fig:lk_vs_lx} we show the absorption-corrected
X-ray luminosity ($L_{X}$) versus the rest-frame $K$-band luminosity ($L_{K}$) of the PL
component for the SMGs which have X-ray observations.  Excepting
2 galaxies with low $L_{X}$ and high $L_{K}$ in the PL component, 
which may host Compton-thick AGN, and 3 galaxies which have
no near-IR PL component and thus likely lack a hot dust component
(similar to the local ULIRGs NGC\,6240 and NGC\,4945), $L_{K}$ of the PL component
is well-correlated with $L_{X}$.  The correlation remains apparent 
even when we restrict the redshift range of the X-ray observed sample
to $2.0 < z < 2.6$, indicating that the correlation is not merely an
artifact of galaxy distance.  The median ratio of the 
$K$-band PL luminosity and $L_{X}$ for all of the SMGs
with X-ray data, $L_{K}(\textrm{PL})/L_{X} = 0.9 \pm 0.4$ with RMS scatter
1.9, is not dissimilar to the $L_{K}/L_{X}$ 
ratio of the average radio-quiet QSO 
SED from \citet{elvis94} ($L_{K}/L_{X} \sim 0.4$), and the range
spans those observed for local PG QSOs, Seyfert 1s, 
and Seyfert 2s \citep[e.g.,][]{alonsoherrero97}.  

The observed $L_{X}-L_{K}$ correlation for our SMGs 
with X-ray data and the similarity of their $L_{K}/L_{X}$ ratios to 
local obscured and unobscured AGN indicate a probable AGN origin of 
the near-IR excess for the galaxies in which the excess is
present in the $K$-band.  To lend credibility to this conclusion we 
also examine the spectroscopic classifications of the SMGs which
have significant near-IR PL contributions.  When we search through 
the rest-frame UV, H$\alpha$, mid-IR, and X-ray spectral classifications
available for the SMGs in our full sample, we find that, of the 25 SMGs
which have fractions of rest-frame $H$-band PL-component luminosity 
greater than 0.2, 20 have been classified as AGN.  The remaining 
5 have starburst classifications, but the only available spectroscopic data is 
from the rest-frame UV, which is very sensitive to dust obscuration
and thus is not a reliable indicator of AGN activity\footnote{As an illustration
of the capability of the rest-frame UV spectra to be misleading, we note
that the two SMGs which do not follow the $L_{X}-L_{K}$ relation and could host 
Compton-thick AGN have starburst classifications 
in the rest-frame UV, but are continuum-dominated in their mid-IR
spectra from IRS.}. Thus, the spectroscopic classifications of the SMGs
in our sample with significant PL components are generally consistent 
with an AGN origin for the near-IR continuum excess in those galaxies.
However, for the SMGs which are at $z<1.8$ and show a near-IR excess
only in one appropriate photometric band, we cannot rule out a
contribution from a 3.3\,$\micron$ PAH emission line without spectroscopic data.
We note that SMGs in our sample without a near-IR excess do not 
universally lack an AGN spectral identification: in fact, $\sim 60$\%
of the galaxies without a near-IR excess have been identified as AGN
through spectroscopy at some wavelength.  However, as we mentioned
in the previous paragraph, some local AGN lack hot dust components;
thus, the presence of AGN in SMGs without a near-IR excess does
not render invalid our conclusion for SMGs with significant excess
near-IR emission. 

Since such a large fraction of our SMG sample ($\sim 70$\%; see \S\ref{sec:excess_desc})
display this near-IR excess which seems to be associated
with AGN (at least in the galaxies with the strongest PL
contribution), it is interesting to determine if the frequency
of near-IR continuum excess observed in our sample
is similar to that of independent SMG samples.  One 
completely independent SMG sample we can compare to is that
catalogued in the Extended-\emph{Chandra} Deep Field-South (E-CDFS)
by \citet{weiss09}.  To reduce the effect of spectroscopic bias from our 
comparison, we examine the number of SMGs in each sample displaying
a near-IR excess relative to the full number of SMGs in the parent sample,
including those whose radio or mid-IR counterparts were not identified.
\citet{wardlow10} find that 13 out of the parent
sample of 126 SMGs in the E-CDFS show a non-stellar excess at 
observed-frame 8.0$\micron$ ($\sim 10$\%), whereas for our sample
44 galaxies out of the parent sample of 150 SMGs in C05 display 
non-stellar excess in their IRAC photometry ($\sim 30$\%).  The large 
difference in frequency of near-IR excess suggests that some bias
towards hot dust emitters and AGN exists in the C05 SMG sample.  Some bias
toward AGN in the C05 sample could have arisen through the spectroscopic
selection, in that SMGs with multiple emission lines in their spectra 
were more likely to be identified than, for example, galaxies with one (or no) emission line. 
However, future submillimeter surveys in additional sky fields are required
to confirm that the C05 SMG sample suffers from an AGN selection bias.

\subsection{Comparison of SMGs to Other High-$z$ Galaxy Populations}

In Paper~I we compared the rest-frame near and mid-IR colors 
of SMGs to the samples of stellar-dominated, rest-frame-UV-selected, 
$z\sim 3$ Lyman-break galaxies (LBGs) and $z\sim 2$ BX/BM galaxies from \citet{reddy06b} 
and the radio-loud, obscured AGN-dominated high-$z$ radio galaxies (HzRGs) 
from \citet{seymour07}, all of which have spectroscopic redshifts.  
Our comparisons indicated the SMGs are brighter and redder than
UV-selected galaxies at high-$z$, suggesting that they have higher
mass, extinction, and possibly hot dust contribution.  SMGs are similarly
bright and red as HzRGs, suggestive of similar mass and
hot dust contribution.  Now we have stellar luminosities and a median stellar mass
for the SMG sample and thus we can compare the stellar components  
of the high-$z$ populations directly.  As in Paper~I, we
prefer to compare only to galaxy samples with spectroscopic redshifts
to eliminate uncertainty associated with redshift; we thus use the
same comparison samples of LBGs, BX/BM galaxies, and HzRGs as in
Paper~I. For the present study we now also include the $z\sim 2$ 
$K$-selected galaxies (distant red galaxies, or DRGs) with spectroscopic redshifts 
from \citet{kriek06,kriek07,kriek08},
to broaden our range of high-$z$ comparison galaxy populations. 
We note that each sample of galaxies representing each population have been
selected by different criteria and at different wavelengths, and thus
each suffers from different selection effects which are beyond the scope of this paper
to analyze. As in Paper I, our comparison is not intended to be 
comprehensive or representative; we merely wish to obtain an idea of
the relationships between well-studied galaxy samples.

Looking at the median rest-frame $H$-band absolute magnitudes
for the various types of high-$z$ galaxies, we find that the 
typical DRG, SMG, and HzRG are $\sim 2$ or more magnitudes brighter
than the typical LBG or BX/BM galaxy.  The reddening-corrected median $M_{H}$ values for
DRGs, SMGs, and HzRGs are $-26.2 \pm 0.1$ (A. Muzzin, private communication),
$-26.0\pm 0.2$, and $-25.7$ \citep{seymour07}, respectively, while the median 
$M_{H}$ for the LBGs and BX/BM galaxies in the \citet{reddy06b}
sample are $-23.8\pm 0.2$ and $-23.6\pm 0.1$, respectively.  
The $\sim 2-2.5$ mag difference between the UV-selected galaxies 
and the DRGs, SMGs, and HzRGs is statistically significant in all cases where
a bootstrapped error in the median is available, and 
corresponds to a factor of $\sim 6-10$ in luminosity.  The
difference in near-IR stellar luminosity probably reflects a 
lower stellar mass in LBGs and BX/BM galaxies,
for three reasons: (1) the hot dust contribution to $M_{H}$ has 
been removed from the SMGs and HzRGs; 
(2) the UV-selected galaxies are less obscured, so 
extinction is unlikely to be a factor in the lower $M_{H}$; and
(3) because we receive UV flux from young stars in UV-selected
galaxies, it is unlikely that the dominant stellar populations in LBGs and BX/BM
galaxies are older (and thus fainter) than in the DRGs, SMGs, and HzRGs. 
The smaller differences in $M_{H}$ among DRGs, SMGs, and HzRGs, however, 
are less clear in origin but are usually not statistically significant.
Hence, we conclude that DRGs, SMGs, and HzRGs have essentially similar
stellar near-IR luminosities, and possibly stellar masses.

Our next step should be to compare the stellar masses of the SMGs, LBGs, 
DRGs, HzRGs, and BX/BM galaxies. Due to the various systematic 
uncertainties associated with $M_{\star}$ estimates
made through fitting stellar population synthesis models, our preference is
to compare galaxy samples whose mass estimates have been derived
in a consistent manner. However, no such concordance exists
in the literature at this moment between the stellar mass-estimating 
methods of the high-$z$ galaxy types selected through different criteria.

In the absence of consistently-derived stellar mass estimates in 
the literature for different high-$z$ populations, we have chosen to
apply our stellar population model plus power law SED fitting technique
to the sample of UV-selected galaxies in \citet{reddy06b} and derive 
stellar masses for that sample in the same manner as we have
done for the SMGs in our sample.  We have chosen to use the LBGs
and BX/BM galaxies in \citet{reddy06b} because complete optical, 
near-IR, and IRAC photometry are readily available 
for a large number of galaxies, along with spectroscopic redshifts.
The median stellar masses we calculate from our method are  
$\langle M_{\star} \rangle = (7.7 \pm 0.2) \times10^{9}\,\msun$ for LBGs
and $\langle M_{\star} \rangle = (5.9 \pm 0.1) \times 10^{9}\,\msun$ for 
the combined BX/BM sample, which are roughly consistent with the stellar
mass estimates of \citet{reddy06b} when scaled to the \citet{kroupa01} IMF.  
When we compare the median stellar mass of SMGs, 
$\langle M_{\star} \rangle = 7.2 \times 10^{10}\,\msun$, to 
the self-consistently calculated median stellar masses of the UV-selected
galaxies, we unambiguously determine that 
the UV-selected galaxies are approximately an order of
magnitude less massive than SMGs.   Such a disparity in mass
does not, however, rule out $z\sim 2$ SMGs as direct evolutionary descendants of 
UV-selected galaxies at $z \sim 3$, because, assuming LBGs contain the necessary
quantity of molecular gas, the time to form the
stellar mass of a typical $z\sim 2$ SMG from the typical mass of 
a typical LBG at the typical LBG star formation rate 
\citep[$\sim 50\,\msun\,\textrm{yr}^{-1}$;][]{kornei10} is $\sim 1$\,Gyr,
roughly the separation in cosmic time between $z\sim 2$ and $z\sim 3$.

\subsection{The Future Evolution of SMGs}\label{sec:gasfrac}

We have now established the typical stellar content of SMGs at $z\sim 2$; yet
we still wish to understand what form the descendants of SMGs will have
in the local Universe.  To explore the future evolution of SMGs after the termination
of the starburst we can employ simple stellar evolution
models based on these star formation time-scales and use the models to predict 
the stellar luminosity distribution of the descendants of SMGs 
from their $M_{H}$ distribution at $z\sim 2$.
For the purposes of our discussion here, since we could derive
no constraints on star formation history from our SED fits 
(see \S\ref{sec:sfhistory}), we conservatively assume a simple
scenario in which we observe the $z\sim 2$ SMGs in the middle of 
last major starburst event in their history; after the burst ends,
they will not undergo subsequent mergers or accretion and the stars
will evolve passively. Our scenario thus provides a lower limit on
the likely $z=0$ luminosity of SMG descendants.

\begin{figure*}
\begin{center}
\includegraphics[scale=0.7,trim=0pt 10pt 0pt 10pt,clip]{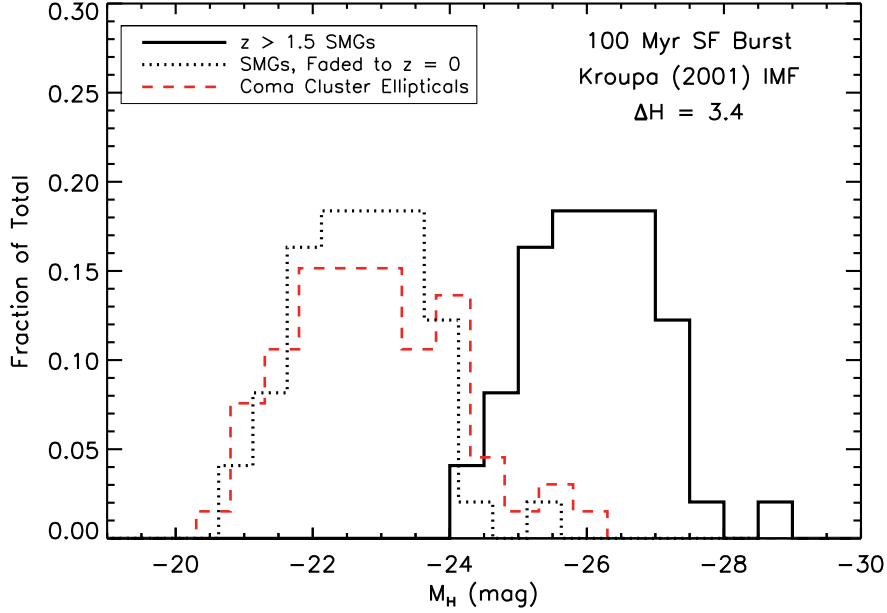}
\caption{Comparison of the absolute $H$ magnitude ($M_{H}$) distribution
of the elliptical galaxies in the Coma cluster to that of the stellar
component of our spectroscopically-identified SMGs at $z>1.5$.  
The distributions match well if we passively evolve the stars in SMGs to 
an age of 10\,Gyr, modeling the SMGs as 100-Myr-long starburst events
which, at $z\sim 2$, we observe halfway through at an age of 50 Myr. 
The match between the faded SMG distribution and the present-day Coma
distribution suggests that, for the predicted time-scale of the starburst
phase in SMGs, SMGs will evolve into galaxies of similar stellar luminosity 
and mass as the massive ellipticals in the Coma cluster if they
evolve passively upon termination of the starburst. }\label{fig:coma_comp_best_fade}
\end{center}
\end{figure*}

\begin{figure*}
\begin{center}
\includegraphics[scale=0.7,trim=0pt 10pt 0pt 5pt,clip]{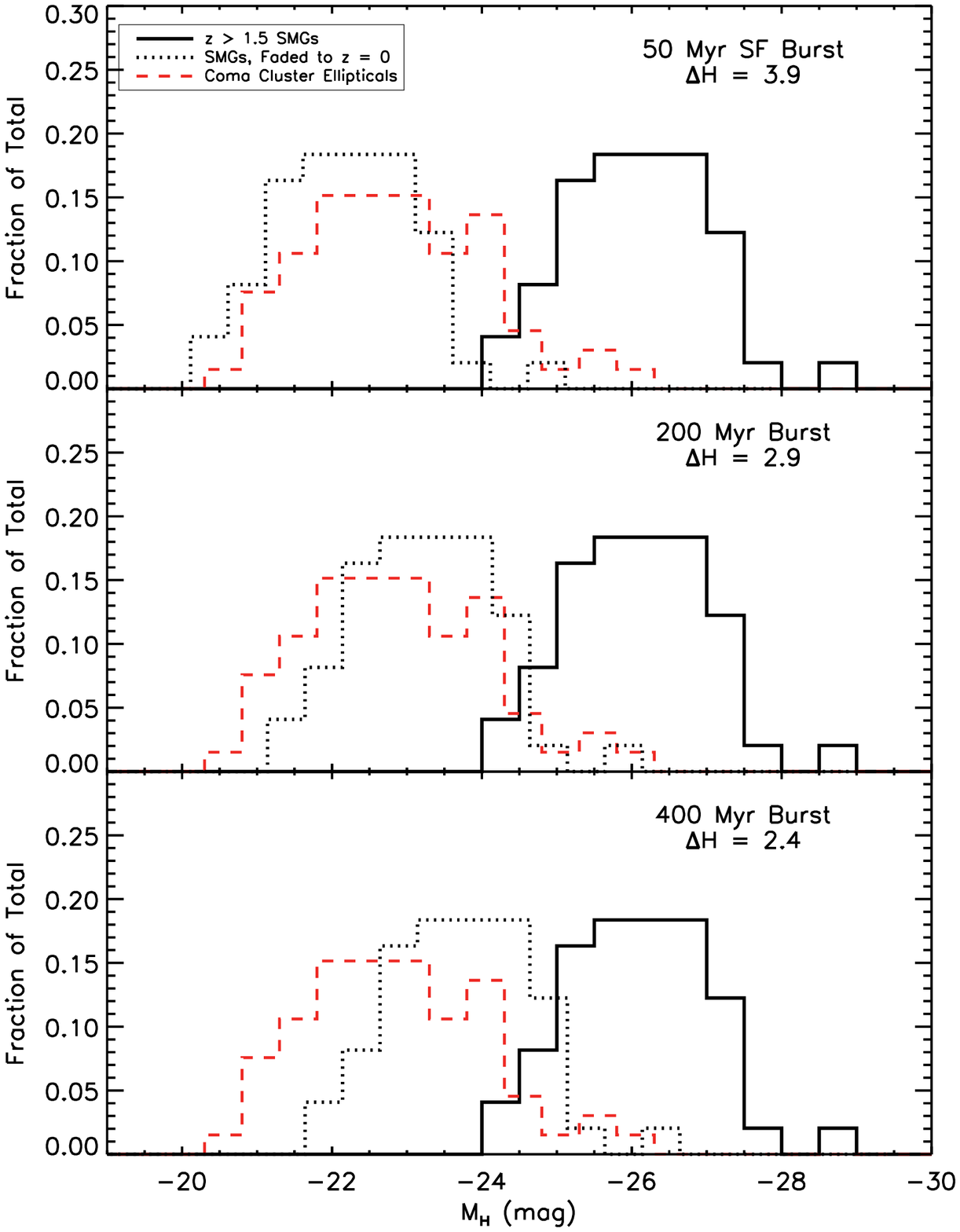}
\caption{Comparison of the absolute $H$ magnitude ($M_{H}$) distribution
of the elliptical galaxies in the Coma cluster to that of the stellar
component of our spectroscopically-identified
SMGs at $z>1.5$, assuming starburst lifetimes for the SMGs
other than 100\,Myr. The fading predicted by modeling SMGs as starbursts
of duration 50 and 200\,Myr cause the $M_{H}$ distribution of SMGs to 
match the Coma cluster similarly as the 100\,Myr burst; however, a 
400\,Myr burst duration will produce SMG descendants which are
more luminous than is typical of the massive ellipticals in 
Coma.}\label{fig:coma_comp_other_fade}
\end{center}
\end{figure*}

We first construct the high-$z$ $M_{H}$ distribution for our sample of $z>1.5$ SMGs
using the power-law continuum-subtracted (stellar) values of $M_{H}$
which have been individually de-reddened and averaged between the IB and CSF star
formation histories. The final
distribution, shown in Figure~\ref{fig:coma_comp_best_fade} as the solid
histogram, has median $\langle M_{H} \rangle = -26.0 \pm 0.2$.  We note
that we do not exclude from the stellar $M_{H}$ distribution the SMGs which have
significant ($> 50$\%) PL components, although their stellar $M_{H}$ values
probably have large uncertainties.  Then, we evolve our sample of $z>1.5$
SMGs to the present day using the M05 stellar population synthesis models
with a \citet{kroupa01} IMF.  We consider as possible scenarios that SMGs are
starbursts of constant SFR of duration 50, 100, 200, and 400\,Myr, covering
the range of estimates of the duration of the starburst in the SMG phase from CO observations
\citep[e.g.,][]{neri03,grevesmg,tacconi06}. We assume
that we observe each SMG halfway though its star formation burst, and then 
compute the fading that occurs between ages 25, 50, 100, and 200\,Myr
and 10\,Gyr (roughly the difference in cosmic time between $z\sim 2$ and $z=0$).
The faded luminosity distributions for each burst duration are shown
as the dotted histogram in Figures~\ref{fig:coma_comp_best_fade} and
\ref{fig:coma_comp_other_fade}, and predict median stellar absolute $H$
magnitudes for the SMGs at $z=0$ of $M_{H} = -22.1, -22.6, -23.1$, and
$-23.6$, respectively, for burst durations of 50, 100, 200, and 400\,Myr.
We observe that the predicted median values of $M_{H}$ are all fainter than 
$L^{\ast}$ at $z=0$ \citep[$M^{\ast}_{H} = -23.7$;][]{cole01}; 
however, recall that these predictions are 
conservative lower limits.  Subsequent mergers or star formation can
increase the mass and luminosity of SMG descendants by varying factors,
and may reasonably be expected.

We can put these aged and faded SMGs into context by comparing them to
a sample of massive elliptical galaxies in the local Universe.
We use as our local massive elliptical galaxy sample the ellipticals
in the Coma cluster since they have a large, homogeneous, near-IR photometry 
data set and optical morphology data set.  We construct the $M_{H}$
distribution for massive ellipticals in Coma using $H$-band 2MASS 
photometry and optical morphologies 
compiled by R.\ Smith and J.\ Lucey (2009, private communication).  
To convert the observed $H$ magnitudes to $M_{H}$, we use a 
distance modulus $m-M = 35.0\,\textrm{mag}$ \citep[e.g.,][]{liu01}.
We include the $M_{H}$ distribution of Coma ellipticals 
in Figures~\ref{fig:coma_comp_best_fade} and \ref{fig:coma_comp_other_fade}
as the dashed histogram.  Somewhat surprisingly, the SMG distributions
faded for 50, 100, and 200\,Myr bursts all reasonably match the 
Coma $M_{H}$ distribution within $\sim 0.5$\,mag: two-sided Kolmogorov-Smirnov
(K-S) tests comparing to Coma distribution to each of the faded SMG distributions
produce probabilities of being drawn from the same parent distribution
of 1.3\%, 35\%, and 13\%, respectively, so we cannot distinguish any of these 
faded distributions from the Coma ellipticals. (We note that the 400 
Myr burst distribution, when compared
to the Coma distribution in the K-S test, produces a probability of 0.07\%.)
As shown in Figure~\ref{fig:coma_comp_best_fade}, the SMG distribution 
faded according to the 100\,Myr burst model appears to 
best match the distribution of the Coma cluster ellipticals, although 
the differences between the 50, 100, and 200\,Myr burst models applied
to the SMG distribution are probably not significant.  
Our comparison of the $M_{H}$ distribution of the elliptical galaxies in
the Coma cluster to faded $M_{H}$ distributions for $z>1.5$ SMGs thus
suggests that, for reasonable starburst durations (50--200\,Myr),
and under the conservative assumption of no significant growth in stellar mass
after the SMG phase terminates,
the $z=0$ SMG descendants may be massive elliptical galaxies 
with a luminosity (and therefore mass) distribution
similar to that of massive ellipticals in the Coma cluster (the 
majority of which fall in the range $0.2L^{\ast}-2L^{\ast}$).  We note that
this range is somewhat lower than the luminosity of present-day SMG descendants
estimated by \citet[][$\gtrsim 3L^{\ast}$]{swinbank06} by comparing the average 
velocity dispersion of 5 SMGs to the present-day Faber-Jackson relation.
In light of the uncertainty in both analyses (arising from star formation history 
assumptions in this paper and the assumptions of virial equilibrium and 
no evolution in velocity dispersion by Swinbank et al.), the two results 
may be reconcilable.  However, our sample of SMGs is $\sim 10$ times larger
than that of \citet{swinbank06}, and therefore is likely more representative of
the range of mass and luminosity among SMGs.

If burst durations of 50--200\,Myr are indeed correct for SMGs, then it is
interesting to determine from them what fraction of today's massive ellipticals 
underwent a SMG phase.  We can use the 
starburst durations which predict the SMG fading that best matches
the Coma ellipticals to estimate the space density
of SMG descendants of similar luminosity/mass to Coma ellipticals, 
and see how that compares to the space density of similar ellipticals
from local galaxy luminosity functions.
We calculate the space density of SMG descendants  
assuming each galaxy goes through 
only one SMG phase using the simple relation
\begin{equation} 
\phi_{D} = \rho_{\textrm{SMG}}\,(t_{\textrm{obs}}/t_{\textrm{burst}}), 
\end{equation}
where $\phi_{D}$ is the comoving space density of the descendants of 
SMGs, $\rho_{\textrm{SMG}}$ is the observed space density of SMGs, 
$t_{\textrm{obs}}$ is the epoch over which we observe SMGs, and 
$t_{\textrm{burst}}$ is the duration of the starburst.  We determine
$t_{\textrm{obs}}$ from the redshift distribution for our SMG sample
\citep[updated from C05 with new results from][]{karin09} after 
correcting for spectroscopic and radio detection 
incompleteness; similar to \citet{swinbank06}, 
we use the $\pm 1\sigma$ interval around
the median of the redshift distribution to define $t_{obs} \simeq 2.3$\,Gyr.  
We estimate the observed space density of SMGs from the integral submm source
counts for $S_{850}>4\,\textrm{mJy}$ found by \citet{coppin06}, 
$844\pm 117\,\textrm{deg}^{-2}$, and the comoving volume between
$z=1.46-3.06$.  In this way, the 50, 100, and 200\,Myr starburst durations
imply that the space density of SMG descendants at
$z=0$ will be $\phi_{D} \sim 2, 1, 0.5 \times 10^{-3}\,\textrm{Mpc}^{-3}$.
When we integrate the $K$-band luminosity function
of early-type galaxies at $z=0$  
of \citet{croton05}, we find that we can match the space density
of SMG descendants of $\phi_{D} \sim 0.5 \times 10^{-3}\,\textrm{Mpc}^{-3}$
by integrating the early-type luminosity function down to $1.3L^{\ast}$.  
For the 100\,Myr-duration burst, we can match the 
predicted space density of $1 \times 10^{-3}\,\textrm{Mpc}^{-3}$  
if we integrate the early-type luminosity function down to $0.8L^{\ast}$.
For the case of the 50\,Myr-duration burst, we can match the 
predicted space density of $2 \times 10^{-3}\,\textrm{Mpc}^{-3}$  
by integrating the early-type luminosity function to $0.4L^{\ast}$.
Thus, our calculations suggest that for starburst durations 
of 50--200\,Myr, SMGs could account for
the formation of the entire population of elliptical galaxies
with luminosities greater than $\sim 0.4L^{\ast} - 1.3L^{\ast}$.  

Given the systematic uncertainty associated with the form of the
stellar IMF for high-$z$ galaxies, our 
median stellar mass for $z>1.5$ SMGs from \S\ref{sec:mstar_results} 
is essentially consistent with $m^{\ast}$, and is consistent with scenarios
which predict that SMGs evolve into massive elliptical galaxies.  Moreover, 
the baryonic masses (gas plus stars) of high-$z$ SMGs, 
which serve as an upper limit on the final stellar mass of SMGs assuming 
no future accretion or gas loss, also point toward a future for SMGs as 
$\sim m^{\ast}$ galaxies.  Assuming the median molecular gas mass of 
CO-observed SMGs (from \S\ref{sec:cocompsec}) is typical of our entire 
sample in an average sense, then the typical 
baryonic mass (gas plus stars) for our SMGs is 
$\langle M_{baryon} \rangle \sim 1.0 \times 10^{11}\,\msun$.
This estimate excludes the contribution from neutral hydrogen; 
if we reasonably estimate that the contribution of neutral gas is similar to that
in the Milky Way and thus equal to the mass of molecular gas,
then the typical baryonic mass of SMGs 
would be $\langle M_{baryon} \rangle \sim 1.3 \times 10^{11}\,\msun$.  Both estimates
of baryonic mass closely resemble the stellar mass of a 
$L^{\ast}$ galaxy in the local universe \citep[$1.4\times 10^{11}\,\msun$;][]{cole01}.
On the other hand, feedback from star formation and/or AGN may result 
in outflows of gas from the galaxy so that only a fraction of
the gas in SMGs at $z\sim 2$ is eventually converted to stars 
\citep[e.g.,][]{begelman85,silk97,dimatteo05,murray10}.  
Nevertheless, the baryonic mass is a strong indicator that the end
product of the starburst phase in a typical SMG will be a galaxy of mass
of order $m^{\ast}$ at $z=0$, within a factor of a few. Even considering the uncertainties, 
the baryonic mass is too small for typical SMGs to represent the formation phase
of the very luminous, rare, cD-type galaxies observed in galaxy clusters; the 
space density of SMGs is also far too high to expect typical SMGs to all be
so massive.  As shown in Figure~\ref{fig:masshist}, a range of approximately
an order of magnitude in stellar luminosity and mass does exist among $z\sim 2$
SMGs, including a high-mass tail, so some SMG descendants will evolve into
massive ellipticals of $3L^{\ast}$ and up as predicted by \citet{swinbank06}. 
Subsequent mergers or gas accretion will also produce more massive descendants.
However, in the absence of significant mass accretion, the average SMG seems 
destined to be an $\sim L^{\ast}$-type galaxy.  

\section{SUMMARY}

In this paper we have fit the rest-frame 
UV through near-IR SEDs of the radio-detected SMGs with spectroscopic
redshifts from \citet{chapman05} to estimate stellar masses. 
For the first time we remove the excess continuum contribution to the 
galaxies' near-IR luminosity to ensure that we derive a mass for the stellar
component only.  Observing that the $K$-band luminosity of this near-IR
excess is well-correlated with hard X-ray luminosity of the galaxy,
we conclude that the near-IR continuum excess is caused by the galaxies'
AGN; however, for only 11\% of our SMGs is the AGN contribution stronger
than the stellar contribution in the rest-frame near-IR.  In trying to fit 
synthetic stellar populations to our continuum excess-subtracted SEDs,
we show that our photometric data 
spanning a broad range of wavelengths are still not sufficient to place
useful constraints on the star formation history of the galaxies and,
thus, the age of the dominant stellar populations, since our galaxies are
heavily reddened and our data are not
well-enough sampled near the strong age indicators of the Balmer and 4000\,\AA\ breaks.  

The median stellar mass of the galaxies in the sample is better 
constrained, however, despite systematic uncertainties.  By taking
a simple, physically motivated approach using
a constant mass-to-light ratio, we find that SMGs have a modest median stellar
mass of $7\times 10^{10}\,\msun$, which is lower than most
previous estimates of stellar mass in SMGs.  Our new, lower stellar
mass estimate for typical SMGs is consistent
with the median dynamical mass of SMGs with CO observations. 
When we compare our typical SMG stellar mass to those
of other high-$z$ galaxy populations, the SMGs appear to
be more massive by a factor of 10 than high-redshift UV-selected galaxies.

We then explore the possible evolution of SMGs after they run out of molecular fuel for 
star formation, based on estimates in the literature for the gas consumption/starburst
timescale in SMGs.  We find that if we passively evolve SMGs to $z=0$,
the range of expected starburst time scales causes the rest-frame $H$ 
absolute magnitude distribution of high-$z$ SMGs to fade so that it nearly
matches the $M_{H}$ distribution of massive elliptical galaxies in 
the Coma galaxy cluster.  The match to the Coma cluster distribution 
suggests that the descendants of typical SMGs will be $\sim m^{\star}$
galaxies similar to Coma ellipticals in the local universe, and is
further supported by the typical baryonic mass of our SMG sample
($\sim 10^{11}\,\msun$) which places an upper limit on the
future stellar mass of SMGs assuming no further cold gas accretion.  Thus, our results confirm the picture 
of high-$z$ SMGs as highly luminous, massive galaxies 
simultaneously experiencing strong, obscured starburst activity and AGN activity 
fueled by massive molecular gas reservoirs, which will evolve into 
$\sim L^{\ast}$-type galaxies following the exhaustion of their molecular gas.

\acknowledgments

We would like to thank the anonymous referee for helpful suggestions
which improved the clarity of the paper.  We thank C.\ Maraston for providing
us with unpublished stellar population synthesis models, as well as C.\ Borys, 
P.\ Capak, and K.\ Bundy for helpful discussions on fitting stellar population
models to observed SEDs.  We thank M.\ Swinbank for helpful discussions as well.
We wish thank R.\ J.\ Smith and J.\ R.\ Lucey for providing their
compilation of morphologies and 2MASS photometry for the Coma Cluster.  We 
acknowledge the use of the NASA Extragalactic Database, as well as E.\ L.\ 
Wright's web-based cosmology calculator.  IRS acknowledges support from STFC.
DMA acknowledges support from the Royal Society and the Leverhulme Trust.


\clearpage

\LongTables

\input{tab1}


\input{tab2}

\input{tab3}
\clearpage

\input{tab4}

\end{document}

%% file: tab1.tex
\begin{deluxetable}{lccccc}
\centering
\tablewidth{0pt}
\tablecolumns{6}
\tablecaption{Photometric Data Used in SED Fits of C05 SMGs\label{tab:smg_fields}}
\tablehead{
\colhead{Field Name} & \colhead{Optical Bands} & \colhead{Near-IR Bands} & 
\colhead{IRAC Bands} & \colhead{$\langle \textrm{\# Optical Bands} \rangle$\tablenotemark{a}} 
& \colhead{Galactic $E(B-V)$\tablenotemark{b}} \\
  \colhead{} & \colhead{} & \colhead{} & \colhead{($\micron$)} & \colhead{(per galaxy)} & \colhead{(mag)}
}
\startdata
CFRS--03h & $U,B,V,R,I$ & $J,K$ & $3.6,4.5,5.8,8.0$ & 6 & 0.071 \\
Lockman & $B,R,I$ & $K$ & $3.6,4.5,5.8,8.0$ & 3 & 0.007 \\
GOODS--N & $U,B,V,R,I,z^{\prime}$ & $J,K$ & $3.6,4.5,5.8,8.0$ & 8 & 0.011 \\
SSA--13 & $B,R,I,z$ & $J,K$ & $3.6,4.5,5.8,8.0$ & 6 & 0.014 \\
CFRS--14h & $B,R,I$ & $J,K$ & $3.6,4.5,5.8,8.0$ & 5 & 0.011 \\
ELAIS--N2 & $B,V,R,I$ & $K$ & $3.6,4.5,5.8,8.0$ & 5 & 0.006 \\
SSA--22 & $B,R,I$ & $J,K$ & $3.6,4.5,5.8,8.0$ & 5 & 0.061 \\
\enddata
\tablenotetext{a}{The median number of optical and near-IR photometric bands per SMG 
actually used in the SED fits.}
\tablenotetext{b}{Estimated line-of-sight reddening due to the Milky Way from \citet{schlegel98}.}
\end{deluxetable}

%% file: tab2.tex
\begin{deluxetable}{lcccccccc}
\centering
\tabletypesize{\tiny}
\tablewidth{0pt}
\tablecolumns{9}
\tablecaption{Results of SED Fitting and Stellar Masses for C05 Spectroscopic
SMG Sample, M05 Models \label{tab:massresults}}
\tablehead{
\colhead{Chapman et al.\ (2005) ID} & \colhead{$z_{spec}$} & \colhead{$M_{H}(\textrm{obs})$\tablenotemark{a}} & 
    \colhead{$M_{H}(\textrm{PL-sub})$\tablenotemark{b}} & \colhead{$\alpha$\tablenotemark{c}} & \colhead{PL Fraction (IB)\tablenotemark{d}}  & 
    \colhead{PL Fraction (CSF)\tablenotemark{e}} & \colhead{$A_{V}$\tablenotemark{f}} & \colhead{$\log{M_{\star}}$\tablenotemark{g}} \\
  \colhead{} & \colhead{} & \colhead{(mag, Vega)} & \colhead{(mag, Vega)} & \colhead{} & \colhead{} & 
     \colhead{} & \colhead{(mag)} & \colhead{($\msun$)}  
}
\startdata
SMM\,J030227.73+000653.5 & 1.408 & $-26.64 \pm 0.16$ & $-26.60 \pm 0.14$ & 3.0 & $0.8_{-0.1}^{+0.1}$ & $0.6_{-0.3}^{+0.1}$ & $1.6 \pm 0.4$ & $11.06 \pm 0.08$ \\
SMM\,J030231.81+001031.3 & 1.316 & $> -22.22$ & $> -22.11$ & 2.0 & $0.0^{+0.9}$ & $0.2_{-0.1}^{+0.7}$ & \nodata & $< 9.26$ \\
SMM\,J030236.15+000817.1 & 2.435 & $-25.91 \pm 0.09$ & $-25.82 \pm 0.09$ & 2.0 & $0.2_{-0.2}^{+0.5}$ & $0.2_{-0.2}^{+0.5}$ & $1.6 \pm 0.4$ & $10.74 \pm 0.05$ \\
SMM\,J030238.62+001106.3 & 0.276\tablenotemark{h} & $-25.68 \pm 0.07$ & $-25.68 \pm 0.07$ & 2.0 & $0.0^{+0.3}$ & $0.0^{+0.4}$ & $1.8 \pm 0.3$ & $10.69 \pm 0.06$ \\
SMM\,J030244.82+000632.3\tablenotemark{i} & 0.176 & $> -22.98$ & $> -22.98$ & 2.0 & $0.0^{+0.1}$ & $0.0^{+0.1}$ & \nodata & $< 9.61$ \\
SMM\,J105151.69+572636.0 & 1.620 & $-27.34 \pm 0.02$ & $-26.83 \pm 0.23$ & 2.0 & $0.5_{-0.5}^{+0.1}$ & $0.2_{-0.2}^{+0.1}$ & $1.0 \pm 1.0$ & $11.16 \pm 0.11$ \\
SMM\,J105155.47+572312.7 & 2.686 & $-25.71 \pm 0.08$ & $-25.13 \pm 0.08$ & 3.0 & $0.8_{-0.1}^{+0.1}$ & $0.7_{-0.1}^{+0.1}$ & $0.3 \pm 0.3$ & $10.47 \pm 0.05$ \\
SMM\,J105158.02+571800.2 & 2.694 & $-27.24 \pm 0.08$ & $-27.24 \pm 0.08$ & 2.0 & $0.0^{+0.2}$ & $0.0^{+0.1}$ & $1.3 \pm 0.4$ & $11.31 \pm 0.05$ \\
SMM\,J105200.22+572420.2 & 0.689 & $-23.88 \pm 0.14$ & $-23.48 \pm 0.08$ & 3.0 & $1.0_{-0.1}$ & $1.0_{-0.1}$ & $1.0 \pm 0.4$ & $9.81 \pm 0.05$ \\
SMM\,J105201.25+572445.7 & 2.148 & $-25.56 \pm 0.04$ & $-24.96 \pm 0.06$ & 3.0 & $0.7_{-0.2}^{+0.1}$ & $0.7_{-0.4}^{+0.1}$ & $1.4 \pm 0.1$ & $10.40 \pm 0.04$ \\
SMM\,J105207.49+571904.0 & 2.689 & $>-26.29$ & $>-26.29$ & 2.0 & $0.0^{+0.6}$ & $0.0^{+0.5}$ & $1.8 \pm 0.3$ & $<10.93$ \\
SMM\,J105219.15+571858.4 & 2.372 & $-26.28 \pm 0.01$ & $-26.28 \pm 0.01$ & 2.0 & $0.0^{+0.4}$ & $0.0^{+0.4}$ & $1.5 \pm 0.0$ & $10.93 \pm 0.05$ \\
SMM\,J105227.58+572512.4 & 2.470 & $-27.08 \pm 0.06$ & $-27.06 \pm 0.04$ & 3.0 & $0.1_{-0.1}^{+0.2}$ & $0.0^{+0.3}$ & $1.8 \pm 0.3$ & $11.24 \pm 0.04$ \\
SMM\,J105227.77+572218.2 & 1.956 & $> -23.32$ & $> -23.11$ & 3.0 & $0.5_{-0.5}^{+0.4}$ & $0.5_{-0.3}^{+0.4}$ & \nodata & $< 9.66$ \\
SMM\,J105230.73+572209.5 & 2.611 & $-27.38 \pm 0.08$ & $-27.31 \pm 0.11$ & 3.0 & $0.1_{-0.1}^{+0.3}$ & $0.3_{-0.3}^{+0.2}$ & $1.0 \pm 0.4$ & $11.34 \pm 0.05$ \\
SMM\,J105238.19+571651.1 & 1.852 & $-25.39 \pm 0.08$ & $-25.38 \pm 0.07$ & 3.0 & $0.1_{-0.1}^{+0.5}$ & $0.0_{-0.0}^{+0.7}$ & $0.4 \pm 0.4$ & $10.57 \pm 0.05$ \\
SMM\,J105238.30+572435.8 & 3.036 & $-27.31 \pm 0.12$ & $-27.31 \pm 0.12$ & 2.0 & $0.0^{+0.1}$ & $0.0^{+0.1}$ & $2.1 \pm 0.3$ & $11.34 \pm 0.08$ \\
SMM\,J123549.44+621536.8 & 2.203 & $-27.13 \pm 0.15$ & $-26.92 \pm 0.21$ & 3.0 & $0.8_{-0.8}^{+0.1}$ & $0.0^{+0.9}$ & $1.0 \pm 0.2$ & $11.20 \pm 0.11$ \\
SMM\,J123553.26+621337.7\tablenotemark{j} & 2.098 & $-25.74 \pm 0.12$ & $-25.31 \pm 0.02$ & 3.0 & $0.7_{-0.2}^{+0.2}$ & $0.8_{-0.4}^{+0.1}$ & $1.2 \pm 0.0$ & $10.54 \pm 0.05$ \\
SMM\,J123555.14+620901.7\tablenotemark{j} & 1.875 & $-27.11 \pm 0.05$ & $-26.76 \pm 0.03$ & 3.0 & $0.8_{-0.1}^{+0.1}$ & $0.7_{-0.5}^{+0.1}$ & $1.8 \pm 0.0$ & $11.12 \pm 0.05$ \\
SMM\,J123600.15+621047.2 & 1.994 & $-26.02 \pm 0.13$ & $-24.69 \pm 0.09$ & 3.0 & $0.9_{-0.1}^{+0.1}$ & $0.9_{-0.9}^{+0.1}$ & $3.0 \pm 0.3$ & $10.30 \pm 0.07$ \\
SMM\,J123606.72+621550.7 & 2.416 & $-25.94 \pm 0.03$ & $-25.01 \pm 0.04$ & 2.0 & $0.8_{-0.1}^{+0.1}$ & $0.8_{-0.2}^{+0.1}$ & $0.3 \pm 0.0$ & $10.42 \pm 0.05$ \\
SMM\,J123606.85+621021.4 & 2.509 & $-26.95 \pm 0.01$ & $-26.95 \pm 0.01$ & 2.0 & $0.0^{+0.3}$ & $0.0^{+0.2}$ & $1.8 \pm 0.0$ & $11.20 \pm 0.05$ \\
SMM\,J123616.15+621513.7 & 2.578 & $-26.91 \pm 0.01$ & $-25.84 \pm 0.73$ & 2.0 & $0.9_{-0.9}^{+0.1}$ & $0.6_{-0.6}^{+0.2}$ & $1.4 \pm 1.4$ & $10.85 \pm 0.27$ \\
SMM\,J123618.33+621550.5 & 2.000 & $-25.85 \pm 0.07$ & $-25.75 \pm 0.07$ & 3.0 & $0.5_{-0.5}^{+0.2}$ & $0.5_{-0.4}^{+0.2}$ & $1.3 \pm 0.4$ & $10.71 \pm 0.05$ \\
SMM\,J123621.27+621708.4 & 1.988 & $-25.77 \pm 0.03$ & $-25.64 \pm 0.08$ & 3.0 & $0.4_{-0.4}^{+0.1}$ & $0.5_{-0.5}^{+0.3}$ & $1.0 \pm 0.4$ & $10.67 \pm 0.05$ \\
SMM\,J123622.65+621629.7 & 2.466 & $-26.84 \pm 0.05$ & $-26.73 \pm 0.10$ & 3.0 & $0.2_{-0.2}^{+0.3}$ & $0.3_{-0.3}^{+0.2}$ & $1.3 \pm 0.4$ & $11.11 \pm 0.05$ \\
SMM\,J123629.13+621045.8 & 1.013 & $-26.42 \pm 0.04$ & $-26.37 \pm 0.04$ & 2.0 & $0.5_{-0.2}^{+0.1}$ & $0.6_{-0.3}^{+0.1}$ & $3.3 \pm 0.3$ & $10.96 \pm 0.04$ \\
SMM\,J123632.61+620800.1\tablenotemark{j} & 1.993 & $-26.30 \pm 0.10$ & $-24.97 \pm 0.09$ & 3.0 & $0.9_{-0.1}^{+0.1}$ & $0.9_{-0.1}^{+0.1}$ & $1.2 \pm 0.3$ & $10.41 \pm 0.07$ \\
SMM\,J123634.51+621241.0 & 1.219 & $-26.29 \pm 0.07$ & $-26.02 \pm 0.06$ & 3.0 & $0.7_{-0.4}^{+0.1}$ & $0.8_{-0.1}^{+0.1}$ & $1.5 \pm 0.3$ & $10.82 \pm 0.04$ \\
SMM\,J123635.59+621424.1\tablenotemark{j} & 2.005 & $-27.90 \pm 0.14$ & $-27.16 \pm 0.19$ & 3.0 & $0.9_{-0.1}^{+0.1}$ & $0.8_{-0.1}^{+0.1}$ & $2.2 \pm 0.5$ & $11.29 \pm 0.10$ \\
SMM\,J123636.75+621156.1 & 0.557 & $-23.54 \pm 0.04$ & $-23.52 \pm 0.05$ & 2.0 & $0.7_{-0.3}^{+0.1}$ & $0.6_{-0.6}^{+0.2}$ & $1.8 \pm 0.0$ & $9.83 \pm 0.06$ \\
SMM\,J123707.21+621408.1 & 2.484 & $-26.93 \pm 0.01$ & $-26.93 \pm 0.01$ & 2.0 & $0.0^{+0.4}$ & $0.0^{+0.3}$ & $2.4 \pm 0.0$ & $11.19 \pm 0.05$ \\
SMM\,J123711.98+621325.7 & 1.992 & $-25.25 \pm 0.07$ & $-25.15 \pm 0.08$ & 3.0 & $0.4_{-0.4}^{+0.3}$ & $0.5_{-0.4}^{+0.2}$ & $1.3 \pm 0.4$ & $10.47 \pm 0.05$ \\
SMM\,J123712.05+621212.3 & 2.914 & $-26.94 \pm 0.07$ & $-26.94 \pm 0.07$ & 2.0 & $0.0^{+0.7}$ & $0.0^{+0.5}$ & $2.5 \pm 0.5$ & $11.19 \pm 0.05$ \\
SMM\,J123721.87+621035.3 & 0.979 & $-25.88 \pm 0.03$ & $-25.80 \pm 0.01$ & 2.0 & $0.6_{-0.2}^{+0.2}$ & $0.5_{-0.4}^{+0.2}$ & $2.1 \pm 0.0$ & $10.74 \pm 0.05$ \\
SMM\,J131201.17+424208.1 & 3.405 & $-27.11 \pm 0.06$ & $-26.37 \pm 0.07$ & 3.0 & $0.5_{-0.3}^{+0.1}$ & $0.5_{-0.5}^{+0.1}$ & $0.6 \pm 0.3$ & $10.96 \pm 0.04$ \\
SMM\,J131208.82+424129.1 & 1.544 & $-26.15 \pm 0.03$ & $-26.11 \pm 0.07$ & 3.0 & $0.5_{-0.3}^{+0.2}$ & $0.0^{+0.3}$ & $2.4 \pm 0.3$ & $10.86 \pm 0.06$ \\
SMM\,J131212.69+424422.5 & 2.805 & $>-25.74$ & $>-25.69$ & 3.0 & $0.2_{-0.2}^{+0.6}$ & $0.0^{+0.6}$ & $1.8 \pm 0.3$ & $<10.70$ \\
SMM\,J131215.27+423900.9\tablenotemark{k} & 2.565 & $-29.49 \pm 0.07$ & $-28.75 \pm 0.03$ & 3.0 & $0.7_{-0.4}^{+0.1}$ & $0.8_{-0.2}^{+0.1}$ & $0.2 \pm 0.2$ & $11.92 \pm 0.04$ \\
SMM\,J131222.35+423814.1 & 2.565 & $-28.32 \pm 0.10$ & $-27.34 \pm 0.18$ & 3.0 & $0.9_{-0.1}^{+0.1}$ & $0.8_{-0.4}^{+0.1}$ & $0.4 \pm 0.4$ & $11.36 \pm 0.10$ \\
SMM\,J131225.20+424344.5 & 1.038 & $-25.73 \pm 0.04$ & $-25.60 \pm 0.04$ & 3.0 & $0.8_{-0.1}^{+0.1}$ & $0.8_{-0.6}^{+0.1}$ & $1.4 \pm 0.1$ & $10.66 \pm 0.04$ \\
SMM\,J131225.73+423941.4 & 1.554 & $-25.25 \pm 0.10$ & $-25.25 \pm 0.10$ & 2.0 & $0.0^{+0.5}$ & $0.0_{-0.0}^{+0.5}$ & $1.0 \pm 0.4$ & $10.52 \pm 0.05$ \\
SMM\,J131228.30+424454.8 & 2.931 & $-26.65 \pm 0.02$ & $-26.14 \pm 0.10$ & 3.0 & $0.6_{-0.2}^{+0.1}$ & $0.7_{-0.2}^{+0.1}$ & $1.2 \pm 0.0$ & $10.87 \pm 0.05$ \\
SMM\,J131232.31+423949.5 & 2.320 & $-27.68 \pm 0.01$ & $-24.31 \pm 0.03$ & 2.0 & $1.0_{-0.1}$ & $1.0_{-0.1}$ & $4.8 \pm 0.0$ & $10.14 \pm 0.05$ \\
SMM\,J131239.14+424155.7 & 2.242 & $-25.44 \pm 0.06$ & $-25.41 \pm 0.09$ & 2.0 & $0.0^{+0.3}$ & $0.2_{-0.2}^{+0.5}$ & $1.3 \pm 0.4$ & $10.58 \pm 0.05$ \\
SMM\,J141741.81+522823.0 & 1.150 & $-28.53 \pm 0.06$ & $-28.12 \pm 0.09$ & 3.0 & $0.9_{-0.1}^{+0.1}$ & $0.9_{-0.1}^{+0.1}$ & $2.8 \pm 0.4$ & $11.67 \pm 0.07$ \\
SMM\,J141742.04+523025.7 & 0.661 & $-25.66 \pm 0.15$ & $-25.49 \pm 0.15$ & 2.0 & $0.8_{-0.1}^{+0.1}$ & $0.8_{-0.1}^{+0.1}$ & $0.6 \pm 0.3$ & $10.62 \pm 0.09$ \\
SMM\,J141750.50+523101.0 & 2.128 & $> -23.65$ & $> -23.65$ & 2.0 & $0.0^{+0.5}$ & $0.0^{+0.3}$ & \nodata & $< 9.88$ \\
SMM\,J141800.40+522820.3 & 1.913 & $-26.67 \pm 0.09$ & $-26.56 \pm 0.10$ & 3.0 & $0.5_{-0.2}^{+0.1}$ & $0.4_{-0.2}^{+0.2}$ & $1.8 \pm 0.3$ & $11.05 \pm 0.07$ \\
SMM\,J141802.87+523011.1 & 2.127 & $-24.30 \pm 0.01$ & $-24.27 \pm 0.02$ & 3.0 & $0.3_{-0.3}^{+0.5}$ & $0.0^{+0.3}$ & $0.2 \pm 0.2$ & $10.12 \pm 0.05$ \\
SMM\,J141809.00+522803.8 & 2.712 & $-27.79 \pm 0.05$ & $-27.79 \pm 0.05$ & 2.0 & $0.0^{+0.1}$ & $0.0^{+0.1}$ & $2.1 \pm 0.3$ & $11.53 \pm 0.04$ \\
SMM\,J141813.54+522923.4 & 3.484 & $> -25.38$ & $> -25.32$ & 2.0 & $0.7_{-0.7}^{+0.2}$ & $0.7_{-0.6}^{+0.1}$ & \nodata & $< 10.55$ \\
SMM\,J163627.94+405811.2 & 3.180 & $>-25.17$ & $>-24.68$ & 2.0 & $0.7_{-0.5}^{+0.1}$ & $0.5_{-0.5}^{+0.2}$ & $0.3 \pm 0.3$ & $<10.30$ \\
SMM\,J163631.47+405546.9 & 2.283 & $-26.19 \pm 0.07$ & $-25.84 \pm 0.09$ & 2.0 & $0.6_{-0.3}^{+0.2}$ & $0.6_{-0.2}^{+0.1}$ & $1.0 \pm 0.4$ & $10.75 \pm 0.05$ \\
SMM\,J163639.01+405635.9 & 1.495 & $-25.74 \pm 0.06$ & $-25.43 \pm 0.11$ & 3.0 & $0.9_{-0.2}^{+0.1}$ & $0.7_{-0.2}^{+0.1}$ & $1.5 \pm 0.6$ & $10.59 \pm 0.07$ \\
SMM\,J163650.43+405734.5 & 2.378 & $-27.07 \pm 0.08$ & $-26.39 \pm 0.28$ & 2.0 & $0.8_{-0.2}^{+0.1}$ & $0.6_{-0.4}^{+0.1}$ & $0.6 \pm 0.3$ & $10.99 \pm 0.13$ \\
SMM\,J163658.19+410523.8 & 2.454 & $-26.61 \pm 0.08$ & $-26.48 \pm 0.09$ & 2.0 & $0.2_{-0.2}^{+0.3}$ & $0.3_{-0.3}^{+0.2}$ & $1.6 \pm 0.4$ & $11.01 \pm 0.05$ \\
SMM\,J163658.78+405728.1 & 1.190 & $-26.22 \pm 0.06$ & $-26.17 \pm 0.08$ & 3.0 & $0.7_{-0.1}^{+0.1}$ & $0.4_{-0.2}^{+0.3}$ & $2.1 \pm 0.3$ & $10.89 \pm 0.06$ \\
SMM\,J163704.34+410530.3 & 0.840 & $-23.57 \pm 0.12$ & $-23.46 \pm 0.23$ & 2.0 & $0.8_{-0.8}^{+0.1}$ & $0.0^{+0.8}$ & $1.3 \pm 0.4$ & $9.81 \pm 0.11$ \\
SMM\,J163706.51+405313.8 & 2.374 & $-27.08 \pm 0.14$ & $-26.76 \pm 0.21$ & 3.0 & $0.7_{-0.2}^{+0.1}$ & $0.4_{-0.3}^{+0.1}$ & $2.1 \pm 0.3$ & $11.13 \pm 0.11$ \\
SMM\,J221733.12+001120.2 & 0.652 & $-25.60 \pm 0.10$ & $-25.53 \pm 0.04$ & 3.0 & $0.6_{-0.2}^{+0.1}$ & $0.7_{-0.2}^{+0.1}$ & $1.8 \pm 0.0$ & $10.63 \pm 0.05$ \\
SMM\,J221733.91+001352.1 & 2.865 & $-26.85 \pm 0.07$ & $-26.49 \pm 0.04$ & 3.0 & $0.6_{-0.2}^{+0.1}$ & $0.3_{-0.3}^{+0.2}$ & $0.9 \pm 0.3$ & $11.01 \pm 0.05$ \\
SMM\,J221735.15+001537.2 & 3.098 & $-25.71 \pm 0.07$ & $-25.22 \pm 0.18$ & 3.0 & $0.3_{-0.3}^{+0.2}$ & $0.4_{-0.3}^{+0.2}$ & $0.8 \pm 0.5$ & $10.51 \pm 0.07$ \\
SMM\,J221735.84+001558.9 & 3.089 & $-26.20 \pm 0.01$ & $-25.79 \pm 0.20$ & 2.0 & $0.6_{-0.4}^{+0.1}$ & $0.3_{-0.3}^{+0.3}$ & $1.2 \pm 0.3$ & $10.74 \pm 0.10$ \\
SMM\,J221804.42+002154.4 & 2.517 & $-26.18 \pm 0.01$ & $-25.92 \pm 0.03$ & 3.0 & $0.6_{-0.6}^{+0.2}$ & $0.5_{-0.5}^{+0.3}$ & $1.2 \pm 0.0$ & $10.79 \pm 0.05$ \\
SMM\,J221806.77+001245.7 & 1.910 & $-26.17 \pm 0.09$ & $-26.17 \pm 0.09$ & 2.0 & $0.0^{+0.1}$ & $0.0^{+0.1}$ & $2.1 \pm 0.3$ & $10.89 \pm 0.07$ \\
\enddata
\tablenotetext{a}{Absolute rest-frame $H$ magnitude derived from observed photometry, averaged between
instantaneous burst and continous SFH model fits.}
\tablenotetext{b}{Absolute rest-frame $H$ magnitude when best-fit PL fraction is subtracted, averaged between
instantaneous burst and continous SFH model fits.}
\tablenotetext{c}{Choice of PL index resulting in best fit for galaxy (lowest $\chi^{2}_{\nu}$).  In the cases
where the best-fit PL fraction is zero, the value listed in this column
indicates the index for which the error bar is calculated.}
\tablenotetext{d}{Best-fit fraction of observed 8.0\,$\micron$ flux contributed by PL component for
instantaneous burst SFH models.}
\tablenotetext{e}{Best-fit fraction of observed 8.0\,$\micron$ flux contributed by PL component for
continuous SFH models.}
\tablenotetext{f}{Average of best-fit $V$-band extinctions from instantaneous burst and continuous SFH fits. The 
listed error is calculated as half the difference between the two averaged values.}
\tablenotetext{g}{Stellar mass derived as described in \S\ref{sec:mass_est_proc}. The 
listed error represents half the difference between the instantaneous burst and continuous SFH values.}
\tablenotetext{h}{SED fits indicate that C05 redshift of $z=0.276$ is incorrect.  Photometric redshift obtained 
from \textsc{hyper-z} suggests alternate redshift of $z=1.81$. $M_{\star}$ has been calculated
using $z=1.81$. Due to uncertainty in the redshift, the results derived
for this galaxy have not been included in the analysis here.}
\tablenotetext{i}{SED fitting fails to converge for any model, possibly due to incorrect galaxy redshift or
error in counterpart selection when measuring optical and IRAC photometry. The results for this galaxy have not 
been included in any of the analysis in the text.}
\tablenotetext{j}{SMG displays UV/blue excess.}
\tablenotetext{k}{Known optical QSO.}

\end{deluxetable}

%% file: tab3.tex
\begin{deluxetable}{lcc}
\centering
\tablewidth{0pt}
\tablecolumns{3}
\tablecaption{Near-IR Continuum Excess-Dominated SMGs\tablenotemark{a} \label{tab:smg_agns}}
\tablehead{
\colhead{Chapman et al.\ (2005) ID} &
   \colhead{\% $L_{H}$ in PL Component (IB)} &
   \colhead{\% $L_{H}$ in PL Component (CSF)}
}
\startdata
SMM\,J123600.15+621047.2$^{b}$ & 67 & 68 \\
SMM\,J123606.72+621550.7 & 56 & 56 \\
SMM\,J123616.15+621513.7 & 71 & 35 \\
SMM\,J123632.61+620800.1$^{c}$ & 67 & 67 \\
SMM\,J131215.27+423900.9 & 45 & 51 \\
SMM\,J131222.35+423814.1 & 60 & 54 \\
SMM\,J131232.31+423949.5 & 97 & 97 \\
SMM\,J163650.43+405734.5 & 53 & 28 \\
\enddata
\tablenotetext{a}{``Continuum-Dominated'' is defined as having greater 
than 50\% of the total $L_{H}$ contributed by the PL component.}
\tablenotetext{b}{Possible Compton-thick AGN; see Fig.\ 4 of
 Alexander et al.\ (2008b).}
\tablenotetext{c}{SMG displays UV/blue excess.}

\end{deluxetable}

%% file: tab4.tex
\begin{deluxetable}{lccccc}
\centering
\tabletypesize{\small}
\tablewidth{0pt}
\tablecolumns{6}
\tablecaption{SMGs with CO Observations\label{tab:co_sources}}
\tablehead{
\colhead{Chapman et al.\ (2005) ID} & \colhead{$z_{spec}$} & \colhead{CO Transition} & 
         \colhead{$M(\hh)$\tablenotemark{a}} & \colhead{$M_{dyn}$} & \colhead{Reference} \\
         \colhead{}                 & \colhead{}          & \colhead{}              & 
         \colhead{$\times 10^{10}\,\msun$}    & \colhead{$\times 10^{11}\,\msun$} & \colhead{}           
}
\startdata
SMM\,J105230.73+572209.5 & 2.611 & 3$\rightarrow$2 & $<1.4$ & \nodata & 1 \\
SMM\,J105238.30+572435.8 & 3.036 & 3$\rightarrow$2 & $<3.0$ & \nodata & 1 \\
SMM\,J123549.44+621536.8 & 2.202 & 3$\rightarrow$2 & 3.2 & 0.94 & 2 \\
SMM\,J123618.33+621550.5 & 2.000 & 4$\rightarrow$3 & 7.2 & 2.2 & 3 \\ 
SMM\,J123634.51+621241.0 & 1.219 & 2$\rightarrow$1 & 6.8 & 7.0 & 4 \\
SMM\,J123707.21+621408.1 & 2.490 & 3$\rightarrow$2 & 2.3 & 2.4 & 2 \\
SMM\,J123711.98+621325.7 & 1.992 & 4$\rightarrow$3 & 3.5 & \nodata & 3 \\
SMM\,J131201.17+424208.1 & 3.405 & 1$\rightarrow$0 & 16 & 1.2 & 1,5 \\
SMM\,J131222.35+423814.1 & 2.565 & 3$\rightarrow$2 & 1.0 & 0.25 & 6 \\
SMM\,J131232.31+423949.5 & 2.320 & 3$\rightarrow$2 & $<2.4$ & \nodata & 1 \\
SMM\,J163631.47+405546.9 & 2.283 & 3$\rightarrow$2 & $<0.9$ & \nodata & 1 \\
SMM\,J163639.01+405635.9 & 1.495 & 2$\rightarrow$1 & $<1.8$ & \nodata & 1 \\
SMM\,J163650.43+405734.5 & 2.380 & 3$\rightarrow$2 & 5.4 & 3.4 & 2 \\
SMM\,J163658.19+410523.8 & 2.454 & 3$\rightarrow$2 & 4.4 & 1.4 & 2 \\
SMM\,J163706.51+405313.8 & 2.374 & 3$\rightarrow$2 & 2.4 & 3.4 & 1 \\
SMM\,J221735.15+001537.2 & 3.098 & 3$\rightarrow$2 & 3.0 & 2.8 & 1 \\
\enddata
\tablecomments{
References: (1) Greve et al.\ (2005); (2) Tacconi et al.\ (2008);
(3) Bothwell et al.\ (2010); (4) Frayer et al.\ (2008); (5) Hainline et al.\ (2006); 
(6) Coppin et al.\ (2008)}
\tablenotetext{a}{$M(\hh)$ calculated assuming a CO-to-$\hh$ conversion factor
$\alpha=0.8$ and $L^{\prime}_{\textrm{CO}(4\rightarrow3)}/L^{\prime}_{\textrm{CO}(1\rightarrow 0)} = 1.0$,  
$L^{\prime}_{\textrm{CO}(3\rightarrow 2)}/L^{\prime}_{\textrm{CO}(1\rightarrow 0)} = 1.0$, or
$L^{\prime}_{\textrm{CO}(2\rightarrow 1)}/L^{\prime}_{\textrm{CO}(1\rightarrow 0)} = 1.0$.}

\end{deluxetable}